\documentclass[aps,prc,twocolumn,superscriptaddress,showkeys]{revtex4-1}

\usepackage{amsmath,amssymb}
\usepackage{graphicx}
\usepackage{multirow}
\usepackage{color}
\usepackage[normalem]{ulem}

\newcommand{\ket}[1]{ | #1 \rangle }
\newcommand{\eq}[1]{\begin{equation} #1 \end{equation}}
\newcommand{\elmx}[3]{\langle #1 | #2 | #3 \rangle}

\makeatletter
\def\p@subsection{}
\makeatother

\begin{document}

\title{Fission barriers of two odd-neutron actinide nuclei taking into account 
	the time-reversal symmetry breaking at the mean-field level}

\author{Meng-Hock Koh}
\affiliation{Department of Physics, Faculty of Science, Universiti
  Teknologi Malaysia, 81310 Johor Bahru, Johor, Malaysia}
\affiliation{University of Bordeaux, CENBG, UMR5797, F-33170 Gradignan, France}
\affiliation{CNRS, IN2P3, CENBG, UMR5797, F-33170 Gradignan, France}

\author{L. Bonneau}
\affiliation{University of Bordeaux, CENBG, UMR5797, F-33170 Gradignan, France}
\affiliation{CNRS, IN2P3, CENBG, UMR5797, F-33170 Gradignan, France}

\author{P. Quentin}
\email[Email address:]{philippe.quentin@tdt.edu.vn}
\affiliation{Division of Nuclear Physics, Ton Duc Thang University, Ho Chi Minh City, Vietnam}
\affiliation{Faculty of Applied Sciences, Ton Duc Thang University, Ho Chi Minh City, Vietnam}
\affiliation{University of Bordeaux, CENBG, UMR5797, F-33170 Gradignan, France}
\affiliation{CNRS, IN2P3, CENBG, UMR5797, F-33170 Gradignan, France}

\author{T. V. Nhan Hao}
\affiliation{Center of Research and Development, Duy Tan University, K7/25 Quang Trung, Danang, Vietnam}
\affiliation{Department of Physics and Astronomy, Texas A\&M University-Commerce, Commerce, TX 75429, USA}
\affiliation{Center for Theoretical and Computational Physics, College of Education, Hue University, 34 Le Loi Street, Hue City, Vietnam}

\author{Husin Wagiran}
\affiliation{Department of Physics, Faculty of Science, Universiti Teknologi Malaysia, 81310 Johor Bahru, Johor}

\date{\today}

\begin{abstract}
\noindent \textbf{Background:} Fission barriers of actinide nuclei have been
mostly and for long been microscopically calculated for even-even
fissioning systems. Calculations in the case of odd nuclei have been
performed merely within a so-called equal-filling approximation (EFA)
–as opposed to an approach taking explicitly into account the time
reversal breaking properties at the mean field level- and for only one
single-particle configuration.\\
\textbf{Purpose:} We study the dependence of the fission
barriers on  various relevant configurations (e.g. to evaluate the
so-called specialization energy). Besides, we want to assess the
relevance as a function of the deformation of the EFA approach which
has been already found out at ground state deformation. \\
\textbf{Methods:} Calculations within the Hartree-Fock plus BCS with
self-consistent particle blocking have been performed using the SkM*
Skyrme effective interaction in the particle-hole channel and a
seniority force in the particle-particle channel. Axial symmetry has
been imposed throughout the whole fission path while the intrinsic
parity symmetry has been allowed to be broken in the outer fission
barrier region. \\
\textbf{Results:} Potential energy curves have been determined for six
different configurations in $^{235}$U and four in $^{239}$Pu. Inner
and outer fission barriers have been calculated along with some
spectroscopic properties in the fission isomeric well. These results
have been compared with available data. The influence of time-reversal
breaking mean fields on the solutions has been investigated. \\ 
\textbf{Conclusions:} A sizeable configuration dependence of the
fission barrier (width and height) has been demonstrated. A reasonable
agreement with available systematic evaluations of fission barrier
heights has been found. The EFA approach has been validated at the
large elongations occurring at the outer barrier region. 
\end{abstract}


\maketitle

\section{Introduction \label{Introduction}}

A microscopic understanding of the nuclear fission process  
remains one of the most complex and challenging problem in
low-energy nuclear physics.  

Although fission-barrier heights are not observable quantities, they
play an important role in determining whether the excited compound
nucleus de-excites through neutron evaporation or fission. They are
also a necessary input for the calculations of fission
cross-sections. From a different point of view, they allow to describe
quantitatively the nuclear stability with respect to spontaneous
fission in competition with other decay modes,
particularly $\alpha$ decay.

Over the years, many microscopic calculations
of the average fission paths of heavy nuclei have been performed,
within mean-field approaches supplemented by the treatment of nuclear
correlations without or with the restoration of some symmetries
spuriously  broken by the mean-field. 
While most of fission-barrier calculations have been
performed for even-mass (with even  proton and neutron numbers) nuclei
(see e.g. \cite{Nikolov_2011,McDonnell_2013,Staszczak_2013,Younes_2009,Warda_2012,Rodriguez-Guzman_2014,Abusara_2010,Lu_2012,Abusara_2012,
  Afanasjev_2013,Hao_2012} for recent related works), there are
comparatively very few microscopic studies dedicated to odd-mass
nuclei and even fewer to odd-odd nuclei. The main reason
is the complication caused by the breaking of time-reversal symmetry
at the mean-field level for a nuclear system involving an odd number
of neutrons and/or protons, considered as identical fermions. 

One of the earlier microscopic study of spectroscopic properties in odd-mass actinides at
very large deformation was performed by Libert and collaborators in
Ref.~\cite{Libert_1980} for the band-head energy spectra in the
fission-isomeric well of $^{239}$Pu within the
rotor-plus-quasi-particle approach. 
More recently, fission-barrier calculations were performed within the Hartree-Fock-Bogoliubov
approach by Goriely \textit{et al.}~\cite{Goriely_2009} for nuclei
with a proton number $Z$ between 88 and 96. 
The resulting fission barriers were then used for the
neutron-induced fission cross-section calculations as part of the
RIPL-3 project published in
Ref.~\cite{Capote_2009}. 
Around the same time, Robledo \textit{et al.} have performed fission-barrier
calculations of the $^{235}$U \cite{Robledo_2009} and $^{239}$Pu
\cite{Iglesia_2009} nuclei, within the equal-filling approximation
(EFA) presented e.g. in Ref.~\cite{Robledo_2008}. In practice, the EFA
consists in occupying pairwise the
lowest single-particle energy levels---exhibiting the two-fold Kramers
degeneracy---and ``splitting'' the unpaired nucleon into two
time-reversal conjugate states with an equal occupation 0.5. In this
way, the time-reversal symmetry is not broken and the calculations are
performed in a way which is very similar to what is done when
describing the ground state of an even-even nucleus.

There are actually two different formalisms in which this EFA is
implemented. One as used in Ref.~\cite{Robledo_2009,Iglesia_2009} deals
with self-consistent calculations of one quasi particle states. It has
been shown to provide the same results as the exact blocking results
within this frame for the time-even part of the
densities~\cite{Schunck_2010}. Another EFA approach will be considered
here in some cases for the sake of comparison with the corresponding
exact calculations which are the
subject of our study. It corresponds here
to an equal-filling approximation to
self-consistent blocked one-particle states.

Although the EFA is likely to be a reasonable approximation, a proper
microscopic description of odd-mass nuclei requires a priori the
consideration of all the effects brought up by the unpaired nucleon. 
This nucleon gives rise to
non-vanishing time-odd densities entering the mean-field
Hamiltonian. The terms involving time-odd densities vanish identically
in the ground-state of even-even nuclei. Their presence for odd-mass
nuclei increases the computing task. As discussed for e.g. in
Refs.~\cite{Quentin_2010,Bonneau_2011}, the time-odd densities cause a
spin polarisation of the even-even core nucleus which results in the
removal of the Kramers degeneracy of the single-particle states. The
recent work of Ref.~\cite{Bonneau_2015} shows that the static magnetic
properties of deformed odd-mass nuclei can be properly described when
taking into account the effect of core polarization
induced by the breaking of the time-reversal symmetry
at the mean-field level. Therefore, it is our purpose here to study
the effect on fission barriers of the time-reversal
symmetry breaking. To do so, we calculate
fission-barrier profiles of odd-mass
nuclei within the self-consistent blocking approach in the 
HF+BCS framework, taking the time-reversal symmetry breaking at the
mean-field level into account.

As well known, some geometrical intrinsic solutions are broken near
both inner and outer barriers. The intrinsic parity is violated for
elongations somewhat before the outer barrier region and 
beyond~\cite{Moller70}. The axial symmetry is also known from a very
long time to be violated in static calculations around the inner
barrier, an effect which is increasing with $Z$ in the actinide region
from, e.g., Thorium isotopes \cite{Pashkevich69}.

Recently it has been suggested that the outer barrier of actinide
nuclei should also correspond to triaxial shapes~\cite{Lu14}. However
the triaxial character of the fission path in both barriers might
vanish or be strongly reduced upon defining it as a least action
trajectory upon making some ansatz on adiabatic mass parameters as well
as on the set of collective variables to be retained. This has been
first discussed in Ref.~\cite{Gherghescu99} for super-heavy nuclei. There,
all quadrupole and octupole (axial and non-axial) degrees of freedom
have been considered. The mass parameters had been calculated
according to the Inglis-Belyaev formula~\cite{Belyaev_1961}. Such a result
has been recently confirmed in non-relativistic~\cite{Sadhukhan14} and
relativistic \cite{Lu14,Zhao16} mean-field calculations. The
calculations of mass parameters have been significantly improved by
using the non-perturbative ATDHFB approach first discussed and used in
Ref.~\cite{Yuldashbaeva_1999}, later revisited in
Ref.~\cite{Baran_2011}. Moreover the intensities of pairing fluctuations
have been included in the set of collective variables together with
the two axial and non-axial quadrupole degrees of
freedom. Calculations in $^{240}$Pu and $^{264}$Fm in
Ref.~\cite{Sadhukhan14} as well as $^{250}$Fm and $^{264}$Fm in
Ref.~\cite{Zhao16} have drawn similar conclusions about the
disappearance or strong quenching of the triaxiality of the fission
paths. These results have been shown to imply very strong consequences
on the spontaneous fission half-lives.

>From these considerations, and keeping in mind the somewhat preliminary
character of our exploration of fission barriers of odd nuclei, we
have deemed as a reasonable first step to stick here to purely axial
microscopic static solutions. 

This paper is organized as follows. In Sec.~\ref{Theoretical framework}, a brief
presentation of the self-consistent blocking HF+BCS formalism and some of its key aspects are
given together with some technical details of the calculations. 
Our results will be presented in Section~\ref{Results: Fission
  barriers} and Section~\ref{Spectroscopic properties in the
  fission-isomeric well}. Finally, the main results are summarised and
some conclusions drawn in the Section~\ref{Conclusion}.

\section{Theoretical framework \label{Theoretical framework}}

The fission-barrier heights have been obtained from
deformation-energy curves whereby the quadrupole
moment has been chosen as the driving coordinate. 
The total energy at specific deformation points 
has been calculated within
the Hartree-Fock-plus-BCS (HF+BCS) approach with blocking, 
and we refer to this as a self-consistent blocking (SCB)
calculation. We will first discuss the details of our SCB calculations
in Subsection~\ref{SCB calculations}, while our approximate treatment
for the restoration of rotational symmetry using the Bohr-Mottelson (BM)
unified model is presented in
Subsection~\ref{Bohr-Mottelson total energy}. 
A detailed discussion about the expressions relating our 
mean-field solutions to the BM model
can be found in Ref.~\cite{Koh_2016}, and  we shall only retain the relevant
expressions herein. Subsection~\ref{Calculation of moment of inertia}
will be devoted to the treatment of the moment of inertia entering the
rotational energy in the BM model, and
  Subsection D to some technical aspects of the calculations.

\subsection{Self-consistent-blocking calculations \label{SCB calculations}}
We assume that the nucleus has an axially symmetrical shape such that 
the projection $\Omega_{k}$ of the total angular momentum onto the
axial symmetry $z$-axis $\hat{j}_z$ of the single-particle state
$\ket{k}$ 
\eq{\elmx{k}{\hat{j}_z}{k} = \Omega_k}
is a good quantum number. The intrinsic left-right (parity) 
symmetry was allowed to be broken around and beyond the top of the
outer-barrier, where such a symmetry breaking is
known to lower the outer-barrier. For our description of
odd-mass nuclei, we have merely considered 
seniority-1 nuclear states in which 
only one single-particle 
state is blocked. The lowest nuclear $K^{\pi}$ state, in general,
corresponds to an unpaired nucleon blocked in the single-particle
state which is the nearest to the Fermi level with
quantum numbers such that $\Omega_k = K$ and, when
parity symmetry is not broken, $\pi_k = \pi$. 
In practice, the blocking procedure translates to setting
the occupation probability $v_k^2$ of the blocked single-particle
state and its pair-conjugate state to 1 and 0,
respectively. 

Such a blocking procedure in an odd-mass nucleus
results in the suppression of the
Kramers degeneracy of the single-particle spectrum. 
As a consequence of time-reversal symmetry breaking at the mean-field level,
the pairs of conjugate single-particle states needed for the BCS pairing treatment 
cannot be pairs of time-reversed states. 
However, without recourse to the Bogoliubov treatment, we were able to unambiguously
identify pair-conjugate states by searching for the maximum overlap
in absolute value between
two eigenstates of the mean-field Hamiltonian,
$\ket{k}$ and $\ket{\widetilde k}$, such that 
$|\elmx{k}{\big( \hat{T}}{\widetilde k}\big)|$, where
$\hat T$ denotes the time-reversal symmetry operator, is as close to 1
as possible. These partner states $\ket{k}$ and $\ket{\widetilde 
k}$ are dubbed as \textit{pseudo-pairs} and they serve as Cooper
pairs in our BCS framework. The value for this overlap will be exactly
1 when time-reversal symmetry is not broken. This procedure for
establishing the BCS pair states when time-reversal symmetry is broken
at the mean-field level has been implemented earlier in the work of
Ref.~\cite{Pototzky_2010}. A more detail discussion can also be found
in Appendix A of Ref.~\cite{Koh_2016}. 

The breaking of the time-reversal symmetry induces terms which are
related to the non-vanishing time-odd local densities in the Skyrme
energy density functionals (see Appendix \ref{Appendix: Skyrme energy density functional}). 
These time-odd local densities are the spin-vector densities
$\mathbf{s}_q$, the spin-vector kinetic energy densities
$\mathbf{T}_q$, the current densities $\mathbf{j}_q$ where the index
$q$ here represents the nucleon charge states. These time-odd local
densities contributes in such a way that the expectation value of the
energy is a time-even quantity as it should.

\subsection{Bohr-Mottelson total energy \label{Bohr-Mottelson total
    energy}}

The total energy within our Bohr-Mottelson approach (see the detailed discussion of Ref.~\cite{Koh_2016}),
is written as
\begin{flalign}
&\elmx{I M K \pi \alpha}{\hat{H}_{{\rm BM}}}{I M K \pi \alpha} \notag \\
&= \elmx{\Psi_{K \pi}^{\alpha}}{\hat{H}_{{\rm eff}}}{\Psi_{K \pi}^{\alpha}} - \frac{1}{2 \: \mathcal{J}} \langle {\rm \bf J}^2 \rangle_{{\rm core}} 
+ \frac{\hbar^2}{2 \: \mathcal{J}} \Big[ I(I+1) \notag \\	
&- K(K-1) 
+ \delta_{K,\frac{1}{2}} a (-1)^{I+\frac{1}{2}} (I+ \frac{1}{2}) \Big]
\end{flalign}
with $\ket{I M K \pi \alpha}$ being the normalized nuclear state
defined by
\begin{flalign}
\ket{I M K \pi \alpha} = & \sqrt{\frac{2I+1}{16 \pi^2}} \Big[
D_{MK}^{I} \ket{\Psi_{K \pi}^{\alpha}} \notag \\ 
& + (-)^{(I+K)} D_{M\, -K}^{I} \hat{T} \ket{\Psi_{K \pi}^{\alpha}} \Big]
\end{flalign}
In the notation above, \textit{I} and \textit{M} are the 
total angular momentum, and its projection on the symmetry axis in the
laboratory frame, respectively. 
The state $\ket{\Psi_{K \pi}^{\alpha}}$ refers to the intrinsic nuclear
state, while $D_{MK}^{I}$ is a Wigner rotation matrix.
The $\langle {\rm \bf J}^2 \rangle_{{\rm core}}$
quantity is the expectation value of the 
total angular momentum operator for a polarized even-even core.
In our model, Coriolis coupling has been
neglected except for the case of $K=1/2$
in which its effect has been accounted for by the
decoupling parameter term. 
The moment of inertia $\mathcal{J}$ and the decoupling
parameter $a$ have been computed from the
microscopic solution of the polarized even-even  
core (see Ref.~\cite{Koh_2016}). 

For the band-head state ($I=K$), the Bohr-Mottelson total energy reduces to
\eq{
E_{K \pi \alpha} = 
\langle \hat{H}_{\rm eff} \rangle - \frac{1}{2 \: \mathcal{J}} \langle
{\rm \bf J}^2 \rangle_{{\rm core}}  
+ \frac{\hbar^2}{2 \: \mathcal{J}} (2 K - \delta_{K,\frac{1}{2}} a)
\label{eq:BM band-head}
}
For given quantum numbers $K$ and $\pi$ (when the intrinsic parity symmetry is present) the
fission-barrier heights have then been
calculated  as
differences of the Bohr-Mottelson energy in 
Eq.~(\ref{eq:BM band-head}) at the
saddle points and the
normally-deformed ground-state $K^{\pi}$ solution.

\subsection{Calculation of the moment of inertia \label{Calculation of moment of inertia}}

Special attention has been paid to the moment of inertia entering
the core rotational energy term given by 
${\rm E_{rot}} = \langle{\hat{\bf J}}^2 \rangle_{{\rm core}}/2 \mathcal{J}$. 
The usual way to handle it is to use the Inglis-Belyaev (IB) formula 
\cite{Belyaev_1961}. It is not satisfactory for at least three reasons.
It is derived within the adiabatic limit of the Routhian
Hartree-Fock-Bogoliubov approach. The Routhian approach is, as well
known, only a semi-quantal prescription to describe the rotation of a
quantal object. Moreover, it is not clear, as we will see, that the
corresponding collective motion is adiabatic. Finally, the IB
formula corresponds to a well-defined approximation to the
Routhian-Hartree-Fock-Bogoliubov approach.

Concerning the last point, as discussed in Ref. \cite{Yuldashbaeva_1999}, the IB moment of
inertia ought to be renormalized to take into account the so-called
Thouless-Valatin corrective terms \cite{Thouless_Valatin_1962} studied in detail in
Ref. \cite{Yuldashbaeva_1999}. 
They arise from the response of the self-consistent
fields with respect to the time-odd density  (as e.g. current and spin
vector densities) generated by the rotation of the nucleus which is
neglected in the IB ansatz. In order to incorporate these
corrective terms in our current approach, the moments of inertia
yielded by the IB formula $\mathcal{J}_{\rm Bel}$ are scaled by
a factor $\alpha$ whose value is taken to be 0.32 following the
prescription of Ref.~\cite{Libert_Girod_Delaroche_1999}: 
\eq{
\mathcal{J}' \; = \; \mathcal{J}_{\rm Bel} \, (1 + \alpha) \,.
}
As a result, one should diminish by the same percentage the rotational
correction evaluated upon using the IB moment of inertia.
Let us remark that the above correction concerns adiabatic regimes of rotation. 

Projecting after variation the $0^+$ state out of a HF+BCS solution,
corresponds, of course, in principle to a better approach to the
determination of the ground-state energy. This has been performed in
Ref.~\cite{Bender_2004} for the fission-barrier of $^{240}$Pu upon
using two Skyrme force parametrizations 
(SLy4 and SLy6 \cite{Chabanat_1998,Chabanat_1998_Erratum}).
These works clearly show that using
the IB approach leads to an overestimation of the rotational
correction by about 10 - 20\% in the region of inner-barrier and
fission-isomeric state and by more than 80\% close to the
outer-barrier. A word of caution on the specific values listed above
should be made, however, since these calculations yield a first $2^+$
energy in the ground-state band which is about twice its experimental
value (83 keV instead of 43 keV).

A third theoretical estimate stems from the consideration of a phenomenological approach belonging to the family of Variable Moment of
Inertia models. It describes the evolution of rotational energies in a
band by consideration of the well known Coriolis Anti-Pairing (CAP)
effect \cite{Mottelson_Valatin_1960} in terms of intrinsic vortical
currents (see e.g. Ref. \cite{Quentin_2004}). The IB treatment to
the moment of inertia corresponds to a global nuclear rotation which
is adiabatic, i.e. corresponding to a low angular velocity $\Omega$, 
or equivalently to a rather small value of the total angular momentum
(also referred to as spin). However, one can compute the average value
of the total angular momentum $I_{\rm av}$ spuriously included in the
mean-field solution as 
\eq{
I_{\rm av}(I_{\rm av}+1) \hbar^2 = \langle \hat{\bf J}^2 \rangle
\label{eq:I average}}
where $\hat{\bf J}$ is the total angular momentum operator,
and find that the value of $I_{\rm av}$ even at ground-state deformation
cannot be considered as small (one finds there that $I_{\rm av}
\approx 13$). Consequently, the moment of inertia entering the
rotational correction term should reflect the fact that the average
$\Omega$ is large. 

Recently, a polynomial expression for the moment of inertia as a
function of $\Omega$ denoted as $\mathcal{J}(\Omega)$ has been
proposed according to this approach to
the Coriolis anti-pairing effect 
(see Ref.~\cite{Quentin_to_be_submitted} and a preliminary account of it in Ref.~\cite{pomorski_2014}). 
This model shall be referred to as the Intrinsic Vorticity Model (IVM) in the
discussion herein. The IVM was found to work well for the rotational
bands in the ground-state deformation  for some actinide nuclei, for
instance a very good agreement for $^{240}$Pu for a value of $I$ as
high as $I_{\rm av} \approx 30$ (where it predicts a rotational energy
differing by only 70 keV from the experimental value). 

Table \ref{table: rotational energy IB, IB+TV and IVM} lists
the spurious rotational energy obtained with 
the IB formula as compared to the IVM rotational energy
for a given value of the total angular momentum $I_{\rm av}$ in the
ground-state of even-even nuclei. 
In all cases, the spurious rotational energy evaluated
with the IB moments of inertia is larger by
about a factor of 2 with respect to the values
obtained in the IVM approach. 
Therefore, the rotational energy 
obtained with the IB formula should
be reduced by approximately 50\%. The same amount of correction is
assumed to apply as well to all other deformations.

\begin{table}
\caption{\label{table: rotational energy IB, IB+TV and IVM}
	Rotational energy (in MeV) calculated from Belyaev formula (IB) and the 
	Intrinsic Vorticity Model (IVM) at the ground-state deformation 
of four even-even nuclei
	as a function of the total angular momentum $I_{av}$
	defined in Eq. (\ref{eq:I average}).}
 \begin{ruledtabular}
 \begin{tabular}{*{6}c}
Nucleus&  $I_{av}$&  &  IB&  &  IVM  \\
\hline
$^{234}$U&  12.988&  &  2.371&  &  1.232	\\
$^{236}$U&  12.905&  &  2.423&  &  1.255	\\
$^{238}$Pu& 13.146&  &  2.441&  &  1.266 	\\
$^{240}$Pu& 13.143&  &  2.408&  &  1.232 	\\
\end{tabular}
\end{ruledtabular}
\end{table}

Incidentally, the 50\% reduction in the rotational energy at all deformation
happens to translate into lowerings
of fission barriers of the same magnitude as those
obtained from the angular momentum projection calculations of
Ref.~\cite{Bender_2004} in $^{240}$Pu. 

One may note that in both the exact or approximate projection formalisms described above, one overlooks -as we will do here- the possible effect of coupling of the pairing mode with the collective shape degrees of freedom, as for instance a possible Coulomb centrifugal stretching (see e.g. Ref. \cite{pomorski_2014}). Indeed, if any, this effect should be more important at the angular momentum value $I_{\rm av}$ than at much lower spins. 

In view of this, we consider to fix ideas,
the following three approaches to the
calculation of the moment of inertia, namely
\begin{itemize}
\item[(i)] the Inglis-Belyaev's formula (IB), 
\item[(ii)] the increase of the
  Inglis-Belyaev moment of inertia by
  32\% (IB+32\%), in order to take into account the
  Thouless-Valatin corrective terms,  
\item[(iii)] the renormalization of the Inglis-Belyaev moment of
  inertia by a factor of 2 (IB+100\%), which
  arises from the 50\% reduction in
  the rotational energy of the intrinsic vorticity model. 
\end{itemize}

\subsection{Total nuclear energies within an approximate projection on good parity states
	\label{Total nuclear energies within an approximate projection on good parity states}}

In the spirit of the unified model description of odd nuclei disentangling the dynamics of an even-even core on
one hand and of the unpaired (odd) nucleon on the other, we factorize
the total wavefunction (with an obvious notation) as
\eq{
\ket{\Psi_{tot}} = \ket{\Phi_{core}} \ket{\phi_{odd}} \,.
}
Similarly we decompose the total Hamiltonian in two separate core and single particle parts
\eq{
\hat{H} = \hat{H}_{core} \hat{h}_{odd} \,.
}
Upon projecting on good parity states both core and odd particle states we get 
\eq{
\ket{\Psi_{tot}} = \ket{\Psi^{+}} + \ket{\Psi^{-}}
}
where the good parity components of $\ket{\Psi_{tot}}$ may be
developed onto core and odd-particle good parity components as 
\eq{
\ket{\Psi ^{+}} = \epsilon \eta \ket{\Phi_{core}^{+}}
\ket{\phi_{odd}^{+}} + \sqrt{1 - \epsilon^2} \sqrt{1 - \eta^2}
\ket{\Phi_{core}^{-}} \ket{\phi_{odd}^{-}}
}
and similarly
\eq{
\ket{\Psi ^{-}} = \epsilon \sqrt{1 - \eta^2} \ket{\Phi_{core}^{+}}
\ket{\phi_{odd}^{-}} + \sqrt{1 - \epsilon^2} \eta^2
\ket{\Phi_{core}^{-}} \ket{\phi_{odd}^{+}}
}
where all kets on the r.h.s. of the two above equations are normalized.
As a result of this, and further making the rough assumption that
$\hat{H}_{core}$ and $\hat{h}_{odd}$ break only slightly the parity,
one gets approximately the energies of the state described by the ket
$\ket{\Psi_{tot}}$ after projection as 
\eq{
E^{+} = \frac{\epsilon^2 \eta^2 (E_{core}^{+} + h_{odd}^{+}) + (1 -
  \epsilon^2) (1 - \eta^2) (E_{core}^{-} + h_{odd}^{-})}{1 -
  (\epsilon^2 + \eta^{2}) + 2 \epsilon^2 \eta^{2}}
}
in the positive parity case and similarly for the negative parity case
\eq{
E^{-} = \frac{\epsilon^2 (1 - \eta^2) (E_{core}^{+} + h_{odd}^{-}) +
  (1 - \epsilon^2) \eta^2 (E_{core}^{-} + h_{odd}^{+})}{(\epsilon^2 +
  \eta^{2}) - 2 \epsilon^2 \eta^{2}}
}
where $E_{core}^{+}$ and $E_{core}^{-}$ are the energies of the
projected core states and $h_{odd}^{+}$ and $h_{odd}^{-}$ are the
diagonal matrix elements
$\elmx{\phi_{odd}^{+}}{\hat{h}_{odd}}{\phi_{odd}^{+}}$ and 
$\elmx{\phi_{odd}^{-}}{\hat{h}_{odd}}{\phi_{odd}^{-}}$.

Only in special cases, can we easily approximate from what we know
about the core projected energies, what are the total projected energy
of the odd nucleus.
  
Let us illustrate the above in two simple cases. The first one is a
favourable one where the odd nucleon has an average parity which is
roughly equal to one in absolute value (e.g.  such that roughly $\eta
= 1$). Then, the total projected energies will be given by 
\eq{
E^{\pi} = E_{core}^{\pi} + e_{odd}
}  
where $e_{odd}$ is the single particle (mean field) energy of the last
nucleon. Now, we recall that the energy of the core state projected
onto a positive parity is lower than (or equal to) what is obtained
when projecting it on a negative parity. Moreover, within the core
plus particle approach, we may approximate (\`{a} la Koopmans) the
total projected nuclear energy $E(K,\pi)$ of the odd nucleus
corresponding to a ($K,\pi$) configuration for the last nucleon as 
\eq{
E(K,\pi) = E_{core}^{+} + e_{odd} = E_{int}(K,\pi) + \Delta
E_{core}^{+}
}
where the intrinsic total energy $ E_{int}(K,\pi)$ results from our
microscopic calculations for the considered single particle ($K,\pi$)
configuration and the corrective energy $\Delta E_{core}^{+}$ is the
gain in energy obtained when projecting the core intrinsic solution on
its positive parity componenet. 

On the contrary whenever the average parity of the odd nucleon state
is close to zero such that roughly $\eta^2 = \frac{1}{2}$ one would
get for instance for the positive parity projected state
\eq{
E^{+} = \frac{\epsilon^2 E_{core}^{+} + (1 - \epsilon^2)
  E_{core}^{-}}{2} + \frac{\epsilon^2  h_{odd}^{+} + (1 - \epsilon^2)
  h_{odd}^{-}}{2} \,,
}
which cannot be simply evaluated without a detailed knowledge of
the projected wave functions.

\subsection{Some technical aspects of the calculations \label{Technical
    aspects of the calculations}}

We have employed the SkM* \cite{SkM*_1982} parametrization as the main
choice of the Skyrme force for our calculations. This Skyrme
parametrization has been fitted to the liquid drop fission-barrier of
$^{240}$Pu and is usually considered as the standard parametrization
for the study of fission-barrier properties, for e.g in
Refs.~\cite{Hao_2012,Bonneau_2004} within the HF framework and
Refs.~\cite{Baran_2015, Schunck_2014, Staszczak_2013} in the
Hartree-Fock-Bogoliubov calculations. 
Two other parametrizations will be also considered here in some
cases, namely the SIII \cite{Beiner_1975} and the SLy5*
\cite{Pastore_2013} parameter sets.

As was done in the study of low-lying band-head spectra in the
ground-state deformation \cite{Koh_2016}, 
to be consistent with the fitting protocol and respect the galilean
invariance, we have neglected the terms 
involving the spin-current tensor density $J_q^{\mu \nu}$
and the spin-kinetic density $\mathbf{T}_q$ by setting the
corresponding coupling constants
$B_{14}$ and $B_{15}$ (see Appendix \ref{Appendix: Skyrme energy density functional} 
for the definition of these constants) to 0 in
the energy-density functional and the Hartree--Fock mean field. 
To make this presentation selfcontained we recall in
Appendix~\ref{Appendix: Skyrme energy density functional}, the
expressions of the Skyrme energy-density functional and the
Hartree--Fock fields, together with the coupling constants in terms of
the Skyrme parameters. In addition, we have also neglected the terms
of the form $\mathbf{s} \cdot \Delta \mathbf{s}$ in the energy-density
functional, where $\bf s$ is the   spin nucleon density, and the
corresponding terms of the Hartree--Fock Hamiltonian. We shall refer
to this as the \textit{minimal time-odd} scheme where only some
combinations of the time-odd densities appearing in the Hamiltonian
density are taken into account. On the other hand, the \textit{full
  time-odd} scheme refers to the case where all time-odd densities are
considered when solving the Hartree--Fock equations. 

The pairing interaction has been approximated with a seniority force 
which assumes the constancy of so-called pairing matrix elements between
all single-particle states belonging to a restricted
valence space. In our case, the valence space has been chosen to
include all single-particle states up to $\lambda_{\rm q} + X$, where
$\lambda_{\rm q}$ is the chemical potential for the charge state $q$
and $X = 6$ MeV. A smoothing factor of Fermi type with a diffuseness
$\mu = 0.2$ MeV (see e.g. Ref.~\cite{Pillet_2002} for details) has
been used to avoid a sudden variation of the single-particle valence
space. The pairing matrix element is given by 
\eq{
g_{\rm q} = \frac{G_{\rm q}}{N_{\rm q} + 11} \,,
}
where $N_{\rm q}$ denotes the nucleon number of charge state $q$. The
pairing strengths $G_{\rm q}$ were obtained by reproducing as best as
possible the experimental mass differences $\Delta_q^{(3)} (N_{\rm
  q})$ of some well-deformed actinide nuclei (for odd $N_{\rm
  q}$-values, see Ref.~\cite{Koh_2016} for further discussions). The
obtained values when using the SkM* parametrization are $G_{\rm n} = G_{\rm p} = 16.0$ MeV.

The calculated single-particle states have been expanded in a
cylindrical harmonic oscillator basis. The expansion needs to be
truncated at some point, and this has been performed according to the
prescription of Ref.~\cite{Flocard_1973} 
\eq{
\hbar \omega_{\bot} \Big( n_{\bot} + 1 \Big) 
+ \hbar \omega_z \Big(n_z + \frac{1}{2} \Big) 
\le \hbar \omega_0 \Big( N_0 + 2 \Big) \,,
}
where the frequencies $\omega_z$ and $\omega_{\bot}$ are related to
the spherical angular frequency, $\omega_0$, by $\omega_0^3 =
\omega_{\bot}^2 \omega_z$. The basis size parameter $N_0 = 14$ which
corresponds to 15 spherical major shells has been chosen. The
two basis size parameters have been optimized for a given Skyrme interaction 
at each deformation point of the neighbouring even-even nuclei while
assuming axial and parity symmetrical nuclear shapes. The optimized
values were then used for the calculations of the odd-mass nuclei.

Numerical integrations were performed using the Gauss--Hermite and
Gauss--Laguerre approximations with 16 and 50 mesh points,
respectively. The Coulomb exchange term has been evaluated
in a usual approximation generally referred to as the Slater approximation~\cite{Slater_1951} even though it had been proposed much earlier by C.F. von Weis{\ae}cker \cite{vonWeizsaecker}.

\section{Fission-barrier calculations \label{Results: Fission barriers}}

\subsection{Fission barriers of odd-mass nuclei without rotational correction 
	\label{Results: Fission-barriers of odd-mass nuclei without rotational correction}}

First the HF+BCS calculations of deformation energy curves as functions of the
quadrupole moment $Q_{20}$, with imposed parity symmetry, were
performed in the two even-even neighboring isotopes of a given
odd-mass nucleus. Subsequently, the calculations for the odd-mass
nucleus were then carried out starting from the converged solutions of
either one of the two even-even neighboring nuclei. It has been
checked that, as it should, the choice of the initial even-even core
solution to be used at a particular deformation point does not affect
the solution of the odd-mass nucleus when self-consistency is achieved. 

For odd-mass nuclei, the choice of the blocked states have been limited to
the low-lying band-head states appearing in the ground-state
well. This corresponds to blocking the single-particle states with
quantum numbers $\Omega^{\pi}$ = 1/2$^{+}$, 5/2$^{+}$, 7/2$^{-}$ and
7/2$^{+}$ for $^{239}$Pu and $^{235}$U, and the additional two
single-particle states with $\Omega^{\pi}$ = 3/2$^{+}$ and 5/2$^{-}$
for $^{235}$U. In all cases, the single-particle state with the
desired $K^{\pi}$ quantum numbers nearest to the Fermi level is
selected as the blocked state at every step of the iteration process. 
However, this selection criterion does not guarantee a converged solution.
There can be, indeed, a problem related to the oscillation of the
blocked state from one iteration to the next. In this case, we were 
forced to perform, instead, two sets of calculations. The blocked
configuration with a lower energy solution was selected as the
solution for the particular $K^{\pi}$ state.

The results of these calculations where intrinsic parity is conserved
are displayed on Figs.~\ref{figure: Pu-239 with parity symmetry
  breaking and no rotational correction} (for $^{239}$Pu) and
\ref{figure: U-235 with parity symmetry breaking and no rotational
  correction} (for $^{235}$U). They lead as well-known, to unduly high
fission barriers for two reasons. One is that a correction for the
spurious rotational energy content (as above discussed and
substantiated below) should be removed throughout the whole
deformation energy curve. The second specific to the outer barrier is
related to the imposition of the intrinsic parity symmetry. This is
why parity-symmetry breaking calculations have been considered. Due to
the huge amount of numerical effort that it involves, we have
considered only some of the lower band-head states in the ground-state
deformation. These are band-head states with $K$ = 1/2, 5/2 and 7/2
states for $^{239}$Pu, and 1/2, 3/2 and two 7/2 states for
$^{235}$U. These parity symmetry breaking calculations were performed
starting from a converged parity-symmetric solution of the respective
$K^{\pi}$ configuration beyond the fission-isomeric well. From this
initial solution corresponding to a given elongation (as measured by
$Q_{20}$), we blocked one single-particle state with $K = \Omega$ and
then performed calculations by constraining the nucleus to a slightly
asymmetrical shape at a finite $Q_{30}$ value for a few
iterations. The constraint on $Q_{30}$ was then released and the
calculations were allowed to reach convergence. Once an asymmetric
solution was obtained, we used it for calculating the next $Q_{20}$
deformation point with an increment of 20 barns. 

\begin{figure*}
\includegraphics[angle=-90,keepaspectratio=true,scale=1.0]{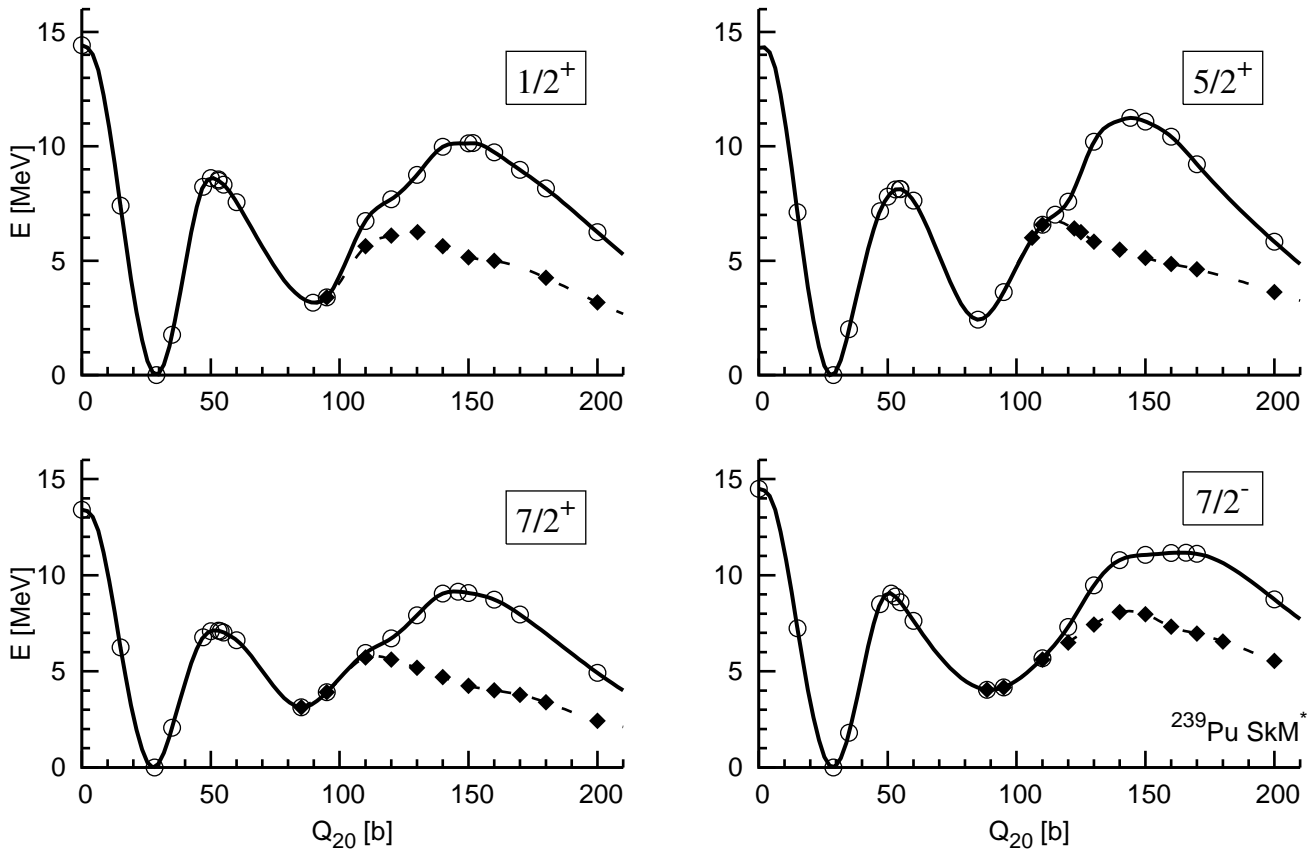}
\caption{Deformation energy curves of
  $^{239}$Pu as functions of $Q_{20}$ calculated with the
  SkM* parametrization and without taking the rotational energy correction into
  account. 
  The $K^{\pi}$ labels refer to the quantum numbers in the parity symmetrical region (unfilled circles). 
  The plotted solutions when this symmetry is broken (filled circles) 
  are obtained by continuity as functions of $Q_{20}$.
  .} 
\label{figure: Pu-239 with parity symmetry breaking and no rotational correction}
\end{figure*}

\begin{figure*}
\includegraphics[angle=-90,keepaspectratio=true,scale=1.0]{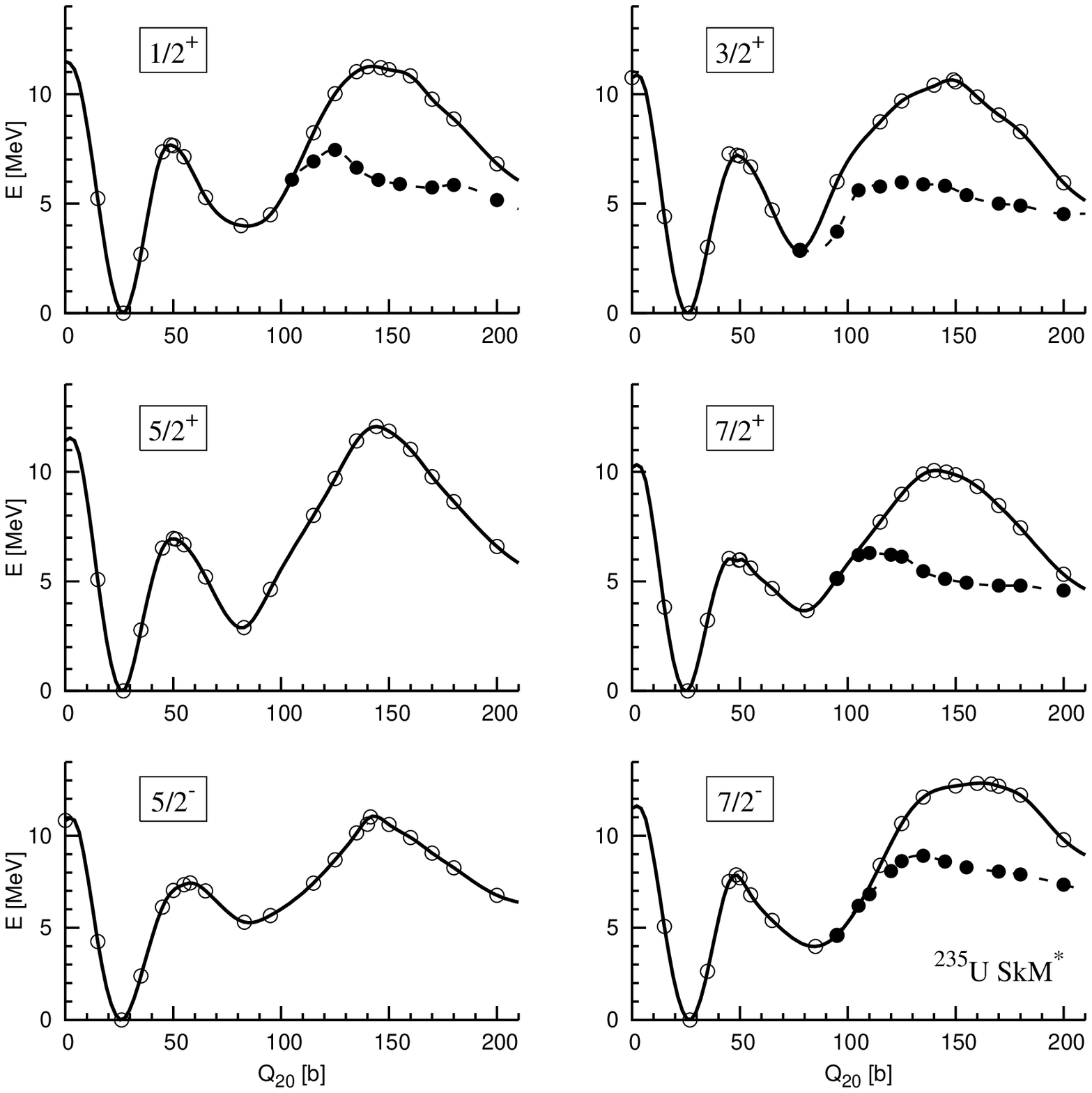}
\caption{Same as Fig.~\ref{figure: Pu-239 with parity symmetry breaking and no rotational correction}
	for $^{235}$U.}
\label{figure: U-235 with parity symmetry breaking and no rotational correction}
\end{figure*}

\begin{figure}
\includegraphics[angle=-90,keepaspectratio=true,scale=0.7]{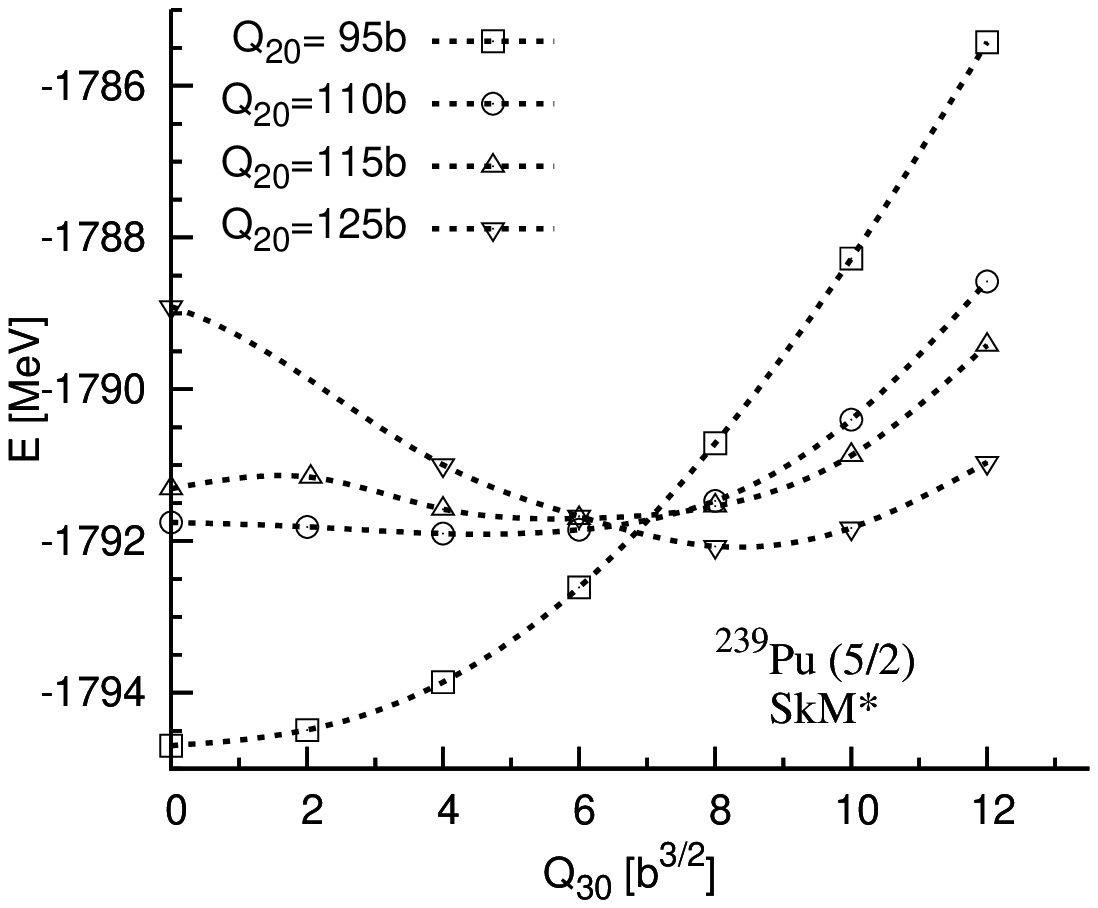}
\caption{Cuts for given values of $Q_{20}$ in the potential-energy surface around the top of the
  outer barrier as a function of the octupole moment $Q_{30}$ (given in
  barns$^{3/2}$) of the 5/2 blocked configuration of $^{239}$Pu. The SkM* parametrization has been used.}
\label{figure: cut in Q20-Q30 plane of 5/2+ Pu-239}
\end{figure}

\begin{figure}[h!]
\includegraphics[angle=-90,keepaspectratio=true,scale=0.7]{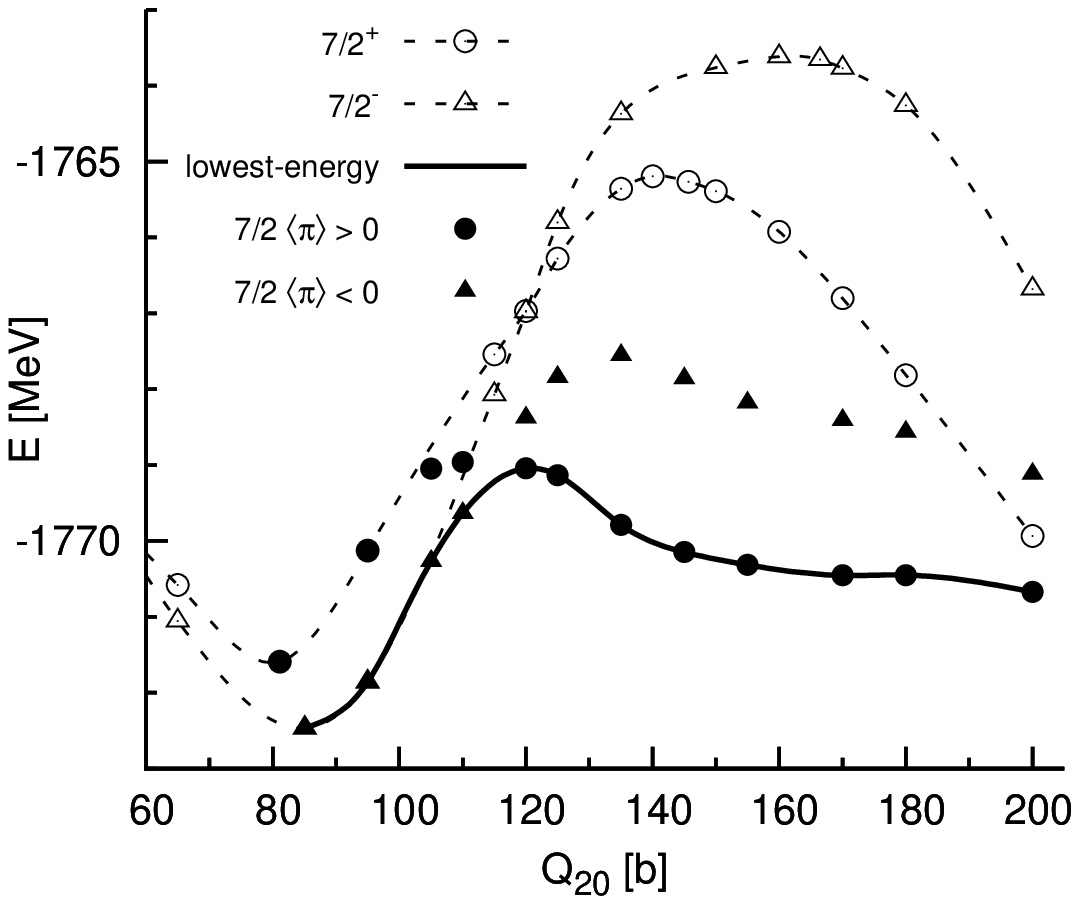}
\caption{A portion of the deformation energy curves of the blocked
  $K=7/2$ configurations in $^{235}$U from the fission-isomeric well
  up to beyond the top of the outer-barrier. The filled symbols refer
  to the local minima as a function of $Q_{30}$ for fixed elongation
  $Q_{20}$ while the unfilled symbols refer to the solutions obtained
  by imposing a left-right symmetry. The solid line connects the
  lowest-energy solutions when the left-right symmetry is broken.
}
\label{figure: parity symmetric and asymmetric deformation energy 7/2 U-235}
\end{figure}

The results of such parity breaking calculations are reported also on 
Figs.~\ref{figure: Pu-239 with parity symmetry breaking and no rotational correction} 
and \ref{figure: U-235 with parity symmetry breaking and no rotational correction}.
Figure~\ref{figure: cut in Q20-Q30 plane of 5/2+ Pu-239} illustrates on a specific example, 
the transition from a symmetrical equilibrium solution at $Q_{20} = 95$ b to 
increasingly asymmetrical equilibrium solutions upon increasing $Q_{20}$. 
At the top of the barrier (corresponding roughly to the $Q_{20} = 110-130$ b range) 
the attained octupole deformations (as measured by $Q_{30}$) reach large values 
which are representative of the most probable fragmentation in the asymmetrical fission mode 
experimentally observed at very low excitation energy in this region.
Of course, upon releasing the symmetry constraint, the parity is no longer a good quantum number. 
Thus, e.g. on Fig.~\ref{figure: Pu-239 with parity symmetry breaking and no rotational correction}, 
the parity-broken energy curve associated with the label $1/2^{+}$ corresponds merely to a $K = 1/2$ solution 
beyond the critical point where the left-right reflection symmetry is lost. 
This may cause some ambiguity in how we define the fission barrier. 
For instance, in the case of $^{235}$U (Fig.~\ref{figure: U-235 with parity symmetry breaking and no rotational correction}) 
we have two $K = 7/2$ solutions of opposite parity. 
On Fig.~\ref{figure: parity symmetric and asymmetric deformation energy 7/2 U-235}, 
we have reported potential energy curves for the two $K = 7/2$ solutions followed by continuity 
upon increasing the deformation from the parity conserved region. 
It turns out that the energy curves of these two solutions are crossing around $Q_{20} = 115$ barns. 
The solution stemming at low $Q_{20}$ from a positive parity configuration becomes energetically favored. 
We could thus define a lowest $K = 7/2$ fission barrier by jumping from one solution to the other. 
Yet, this overlooks two problems. 
One which will be touched upon below, is the projection on good parity states. 
The other is the fact that we do not allow here for a residual interaction between the two configurations, 
a refinement that is beyond the scope of our current approach.

As expected, the parity-symmetry-breaking calculations do yield a
substantial effect on the intrinsic deformation energies around the outer
fission-barrier. Its height for the 1/2 configuration in $^{239}$Pu is
lowered by about 3.9 MeV with respect to the symmetrical barrier,
leading to a calculated height $E_B = 6.3$~MeV. The outer-barrier
height for the 5/2 configuration, in the same nucleus, was found to be
$E_B = 6.6$~MeV, corresponding to an even larger reduction of 4.7 MeV
with respect to the left-right symmetric barrier height. Important
reductions of fission barrier heights are also obtained in the
$^{235}$U case (see Fig.~\ref{figure: U-235 with parity symmetry
  breaking and no rotational correction}). One lowers the $K = 1/2$
outer barrier by 3.7 MeV and by 5.4 MeV in the $K = 7/2$ case. 

Associated with this substantial gain in energy upon releasing the
left-right reflection symmetry, one observes also a very important
lowering of the elongation at the outer fission saddle point,
resulting in a reduced barrier width and therefore in a strong further
enhancement of the barrier penetrability.

To generate relevant outer barrier heights, one has in principle to
project our solutions on good parity states. In the absence of such
calculations for the odd nuclei under consideration here, one may
propose some reasonable estimates taking stock of what we know about
the projection of a neighboring even-even core nucleus. As discussed
in Subsection~\ref{Theoretical framework}.\ref{Total nuclear energies
  within an approximate projection on good parity states} however,
this is only possible whenever the single-particle wavefunction of the
last (unpaired) nucleon corresponds to an average value of the parity
operator which is close to 1 in absolute value. This is not always the
case as exemplified on Fig.~\ref{figure: evolution of 7/2 s.p. level
  in U-235} corresponding to two low excitation energy $ K = 7/2$
configurations in the $^{235}$U nucleus. They are followed, as we have
already seen, by continuity from slightly before the isomeric-fission
well to much beyond the outer barrier. One of these two solutions
stemming from a $K^{\pi} = 7/2^{-}$ configuration at small elongation
keeps up to $Q_{20} = 120-130$ b an average parity reasonably close to
1. On the contrary, the other $ K = 7/2$ solution involves in the
outer barrier region, a large mixing of contributions from both
parities. We will therefore be only able to evaluate the fission
barrier of the former and will not propose any outer fission barrier
height for the latter. 

\begin{figure}[h!]
\includegraphics[angle=-90,keepaspectratio=true,scale=0.75]{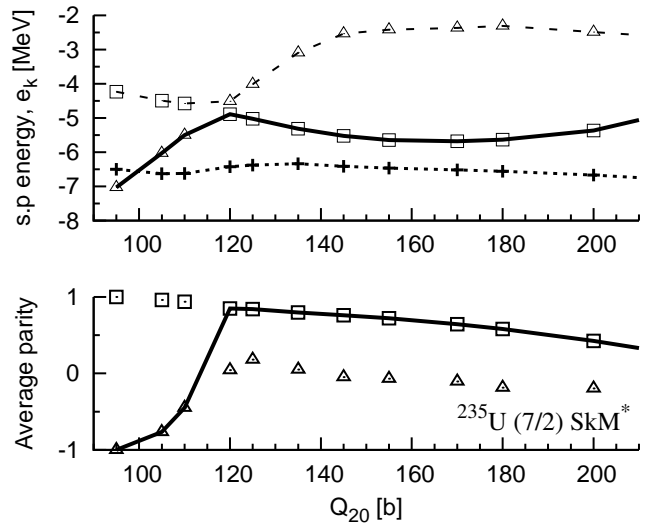}
\caption{(Top): Evolution of the single-particle energies for two $\Omega$ = 7/2 states near the BCS chemical potential (marked with crosses)
	as functions of $Q_{20}$ obtained in the parity asymmetric calculations 
	of $^{235}$U. The solid line connects the blocked single-particle states as a function of deformation.
	(Bottom): Average parity of the above considered blocked single-particle states as a function of $Q_{20}$.} 
\label{figure: evolution of 7/2 s.p. level in U-235}
\end{figure}

In the work of Ref.~\cite{Hao_2012} one has described the fission barrier of
$^{240}$Pu nucleus within the Highly Truncated Diagonalization
approach, to account for pairing correlations while preserving the
particle-number symmetry. Such solutions have been projected on good
parity states after variation. The parity-projection calculation had
no effect on the total binding energy at the top of the outer
fission-barrier, where the value of $Q_{30}$ was found to be very large. 
In contrast, projecting on a positive parity state causes a lowering
of the total binding energy in the fission-isomeric well. 

Using the notation of Subsection~\ref{Theoretical
  framework}.\ref{Total nuclear energies within an approximate
  projection on good parity states}, one has obtained in
Ref.~\cite{Hao_2012} for the $^{240}$Pu nucleus, a positive correcting
energy $\Delta_{core}^{+}$ about equal to 350 keV for the
fission-isomeric state and which vanishes at the top of the outer
barrier.

According to the discussion of Subsection~\ref{Theoretical
  framework}.\ref{Total nuclear energies within an approximate
  projection on good parity states}, out of all the configurations
considered up to the isomeric state in $^{235}$U and in $^{239}$Pu,
only the $ K^{\pi} = 1/2^{+}$ and $7/2^{+}$ configurations in $^{235}$U,
and the $7/2^{+}$ configuration in $^{239}$Pu qualify to allow us to
propose reasonable estimates of the outer fission barrier heights (see
Table~\ref{table: average parity}). 

\begin{table}[h]
\caption{\label{table: average parity} 
Expectation value of the parity of the parity operator for the blocked
single-particle state nearest to the Fermi level in both considered
odd nuclei corresponding to a specific K configuration in the $Q_{20}
= 120-130$~barn region. The SkM*interaction has been used. Only the
lowest energy solutions have been considered for a given $K$ value. In
the single case $K = 7/2$ where two solutions stemming by continuity
from states with the same $K$ and opposite $\pi$ values were close
enough in energy (a couple of MeV) we have reported the parity
expectation value of both, putting in parenthesis the solution with
the higher energy.}
\begin{ruledtabular}
\begin{tabular}{*{5}c}
$K$ & 1/2 & 3/2 & 5/2 & 7/2 \\
\hline
$^{235}$U & 0.76 & $-0.53$ & 0.06 & 0.85 (0.10) \\
$^{239}$Pu & -- & $-0.13$ & -- & 0.83 (0.19)
\end{tabular}
\end{ruledtabular}
\end{table}

\subsection{Inclusion of rotational energy and sensitivity of
  fission-barrier heights to the moment of inertia 
	\label{Sensitivity to moment of inertia}}

Table~\ref{table: sensitivity of fission-barrier heights all configurations}
displays the 
inner-barrier height $E_A$, the fission-isomeric energy $E_{\rm IS}$ 
and the outer-barrier height $E_B$, obtained within the Bohr-Mottelson
unified model (therefore including the rotational energy). 
Parity symmetric and asymmetric (when available) outer-barrier
heights are both tabulated for completeness. It should be emphasized that
the notation $E_{\rm IS}$ used here is not synonymous with the usual
meaning of fission-isomeric energy often denoted by $E_{\rm II}$. The
latter refers to the energy difference between the lowest-energy
solutions in the fission-isomeric and
ground-state wells. The corresponding 
results will be reported in Section 
\ref{Spectroscopic properties in the fission-isomeric well}, 
while the former is the energy difference between
given $K^{\pi}$ quantum numbers in the two wells.

 \begin{table*}
 \caption{\label{table: sensitivity of fission-barrier heights all configurations} 
Inner-barrier height $E_A$, fission-isomeric energy $E_{\rm
  IS}$ (with respect to the ground-state solution), 
and outer-barrier height $E_B$ for the two considered odd-neutron nuclei.
The SkM* parametrization has been used. Three values (in MeV) were given in all cases,
corresponding to different prescription for the moments of inertia (see the discussion in Section~\ref{Theoretical framework}).}
 \begin{ruledtabular}
 \begin{tabular}{*{17}c}
\multirow{2}{*}{Nucleus}&  \multirow{2}{*}{$K^{\pi}$}&  \multicolumn{3}{c}{$E_A$}&  &  
	\multicolumn{3}{c}{$E_{\rm IS}$}&  &  
	\multicolumn{3}{c}{$E_{B}$ (symmetric)}&  &  \multicolumn{3}{c}{$E_{B}$ (asymmetric)} \\
\cline{3-5} \cline{7-9} \cline{11-13} \cline{15-17}
	&	&	IB&  IB+32\% &  IB+100\% &  &  IB&  IB+32\% &  IB+100\% &  &  IB&  IB+32\% &  IB+100\% &  &  IB&  IB+32\% &  IB+100\% \\
\hline
\multirow{6}{*}{$^{235}$U}
&  1/2$^+$&  6.57&  6.83&  7.11&  &  2.60&  2.94&  3.30&  &  8.60&  9.23&  9.90&  &  5.31&  5.83&  6.38	\\
&  3/2$^+$&  6.19&  6.43&  6.69&  &  1.48&  1.81&  2.16&  &  8.12&  8.72&  9.37&  &  \\	
&  5/2$^+$&  5.83&  6.09&  6.37&  &  1.44&  1.78&  2.15&  &  9.57&  10.17& 10.80& &  \\	
&  5/2$^-$&  6.32&  6.59&  6.87&  &  3.97&  4.28&  4.62&  &  8.21&  8.81&  9.46&  &  \\	
&  7/2$^-$&  6.97&  7.18&  7.41&  &  2.70&  3.00&  3.32&  &  10.25& 10.85& 11.49& &  \\	
&  7/2$^+$&  4.75&  5.04&  5.35&  &  2.21&  2.55&  2.91&  &  7.29&  7.93&  8.61&  &  4.03&  4.54&  5.09	\\
\hline
\multirow{4}{*}{$^{239}$Pu}
&  1/2$^+$&  7.43&  7.71&  7.98&  &  1.70&  2.05&  2.43&  &  7.63&  8.24&  8.88&  &  \\	
&  5/2$^+$&  6.97&  7.25&  7.54&  &  0.96&  1.30&  1.67&  &  8.83&  9.40&  10.00& &  \\ 
&  7/2$^-$&  8.10&  8.32&  8.56&  &  2.74&  3.05&  3.37&  &  8.75&  9.32&  9.93&  &  \\ 
&  7/2$^+$&  5.90&  6.18&  6.48&  &  1.72&  2.05&  2.40&  &  6.63&  7.22&  7.86&  &  3.80&  4.25&  4.72 \\
 \end{tabular}
 \end{ruledtabular}
 \end{table*}

It can be seen from Table~\ref{table: sensitivity of fission-barrier heights all configurations} 
that the rotational-energy correction calculated
using the Inglis-Belyaev formula gives too low an outer fission barrier in
some cases, as compared to the empirical values found to be 
within the range of 5.5 to 6.0 MeV (see 
Table~\ref{table: fission-barrier heights comparison for odd-mass
  nuclei} presented in the next Subsection).
The increase in the IB moments of inertia by 32\% and 100\% as discussed in Section~\ref{Theoretical framework}
results in an increase, on the average, of $E_A$ and $E_{IS}$ by about 0.27 MeV and 0.35 MeV, respectively
while the parity symmetric $E_B$ increases by about 0.64 MeV. 

Among the three different considered energy differences, $E_B$  is
found to be the most sensitive one to the variation of the moment of
inertia as expected in view of the well-known increase of the
rotational energy correction with the elongation. 

\subsection{Comparison with empirical values and other calculations}

Before comparing our fission-barrier heights to other available data, 
some corrections should be made.
The corrections considered herein, stem from approximations of different nature:
the so-called Slater approximation to the Coulomb exchange
interaction, the truncation of the harmonic-oscillator basis, 
and the effect of triaxiality around the inner-barrier ignored here.

We shall discuss first the corrections to be made for the inner-barrier heights.
A test study on the impact of basis size parameter on the fission-barrier heights
is presented in 
Appendix~\ref{Appendix: Effect of basis size on fission-barrier heights}. 
As discussed therein, the inner-barrier height is estimated to be
lowered by about 300 keV when increasing the basis size parameter $\rm
N_0$ to a value where this relative energy may be considered to have converged. 
Moreover the use of Slater approximation was found
in Ref.~\cite{Bloas_2011} to underestimate the
inner-barrier height of $^{238}$U by about also 300 keV. 
Assuming that a similar correction
applies to the two considered nuclei irrespective of the 
$K^{\pi}$ quantum numbers, our inner-barrier height should be
increased by the same magnitude.

 \begin{table*}[t]
 \caption{\label{table: fission-barrier heights comparison for odd-mass nuclei} 
	Comparison between various estimates of the inner $E_A$ and outer-barrier $E_B$ heights (given in MeV)
	of the two considered odd-neutron nuclei.
	Our calculated fission-barrier heights corresponding to the experimental $K^{\pi}$ quantum numbers,
	at ground state deformation,
	are listed in the last column, 
	whereby these values have been obtained after taking the various corrections into account.
	}
 \begin{ruledtabular}
\begin{tabular}{*{19}c}
\multirow{2}{*}{Nucleus}&  \multirow{2}{*}{$K$}&  \multicolumn{2}{c}{Ref. \cite{Robledo_2009,Iglesia_2009}}&  &
	\multicolumn{2}{c}{Ref. \cite{Moller_2009}}&  &	
	\multicolumn{2}{c}{Ref. \cite{Goriely_2009}}& &  
	\multicolumn{2}{c}{Ref. \cite{Smirenkin_1993}}&  &  
	\multicolumn{2}{c}{Ref. \cite{Bjornholm_1980}}&  &
	\multicolumn{2}{c}{present work}	\\
\cline{3-4}  \cline{6-7} \cline{9-10} \cline{12-13} \cline{15-16}  \cline{18-19}
&	&	$E_A$&  $E_B$&  &	$E_A$&  $E_B$&  &
	$E_A$&  $E_B$&  &  $E_A$&  $E_B$&  &  $E_A$&  $E_B$&  &  $E_A$&  $E_B$	\\
\hline 
\multirow{6}{*}{$^{235}$U}
&  	1/2$^+$&	9.0&  8.0& &	
	\multirow{6}{*}{4.20}&  \multirow{6}{*}{4.87}&	&
	\multirow{6}{*}{5.54}&	\multirow{6}{*}{5.80}&  &
	\multirow{6}{*}{5.25}&  \multirow{6}{*}{6.00}&  &
	\multirow{6}{*}{5.9}&  \multirow{6}{*}{5.6}&	&
	5.81&  6.18	\\
&  	3/2$^+$&	-&  -& &&	&&&&&&&&&&&	5.39&	-	\\
&  	5/2$^+$&	-&  -& &&	&&&&&&&&&&&	5.07&	-	\\
&  	5/2$^-$&	-&  -& &&	&&&&&&&&&&&	5.57&	-	\\
&	7/2$^-$&	\multirow{2}{*}{8.5}&  \multirow{2}{*}{7.2}& &	&   &&&&&&&&&&& 6.11&  -				\\
&	7/2$^+$&			&			&	&	&	&&&&&&&&&&&	4.05&  4.89 		\\
\hline
\multirow{4}{*}{$^{239}$Pu}
&  	1/2$^+$& 	11.0&  8.5& &	
	\multirow{4}{*}{5.73}&	\multirow{4}{*}{4.65}&	&
	\multirow{4}{*}{5.96}&	\multirow{4}{*}{5.86}&  &
	\multirow{4}{*}{6.20}&	\multirow{4}{*}{5.70}&  &
	\multirow{4}{*}{6.2}&	\multirow{4}{*}{5.5}&	&
	6.68&	-	\\
&	5/2$^+$&	11.5&  9.0& &	&	&&&&&&&&&&& 6.24&  -	\\
&	7/2$^-$&	\multirow{2}{*}{11.0}&  \multirow{2}{*}{8.5}&   &	&	&&&&&&&&&&&	7.26&  -	\\
&	7/2$^+$&			&			&	&	&	&&&&&&&&&&&	5.18&  4.52	\\
\end{tabular}
 \end{ruledtabular}
 \end{table*}

Let us consider the impact of breaking the axial symmetry should
around the top of the inner barrier. When breaking this symmetry, $K$
is no longer a good quantum number and this may pose a problem in the
blocking procedure for an odd-mass nucleus since the single-particle
states will contain to some extent mixtures of $K$ quantum number
components. As a simple ansatz, overlooking these potential
difficulties, we estimate the lowering of the inner barrier of
odd-mass nuclei by using the results obtained  in similar triaxial
calculations for even-mass nuclei, taking stock of the results of
Ref.~\cite{Algerian_2016} where the same SkM* parametrization and
seniority residual interaction have been used. Thus assuming that the
effect of including the triaxiality is the same as in $^{236}$U (for
$^{235}$U) and as in $^{240}$Pu (for $^{239}$Pu) for all considered
blocked configurations, we expect a reduction in the inner-barrier
height by about 1.3 MeV. 

Taking the three above mentioned corrections into account,
we obtain a total reduction of the inner-barrier height by about 1.3 MeV.

Next, we consider the isomeric energies $E_{IS}$. We estimate that
the finite basis size effect (see Appendix~\ref{Appendix: Effect of
  basis size on fission-barrier heights}) results in an overestimation
of this energy by about 0.5 MeV. The exact Coulomb exchange
calculations of Ref.~\cite{Bloas_2011} have shown that the Slater
approximation yielded an underestimation of the isomeric energy of
$^{238}$U of about 0.3~MeV.

As for the outer barrier now, exact Coulomb exchange calculations have
not been performed---due to corresponding very large computing
times---for these very elongated shapes in this region of nuclei. As
discussed in Ref.~\cite{Bloas_2011} most of the correction comes from
an error in estimating the Coulomb exchange contributions in low
single-particle level density regimes Therefore as far as $E_B$ is
concerned, we assume that this correction depends only on the
treatment of the ground-state and therefore should be the same as 
what was obtained for $E_A$, namely an underestimation of 0.3~MeV.
The finite basis size effect, as evaluated in a particular case in
Appendix~\ref{Appendix: Effect of basis size on fission-barrier
  heights} corresponds to an overestimation of about 0.5 MeV.
The nett effect of the corrective terms for the outer-barrier height is
therefore a decrease by about 0.2~MeV.

When including all the above
corrections and using the doubled moment of inertia (IB+100\% scheme),
we obtain inner-barrier heights for the
different blocked configurations ranging from 5.0 to 6.2
MeV for $^{235}$U, and from 5.1 to 7.3 MeV for $^{239}$Pu.
The left-right asymmetric outer-barrier heights 
lie within the range of 4.8 to 6.2~MeV for the $^{235}$U
nucleus, and 4.5~MeV for 7/2$^+$ configuration in the $^{239}$Pu nucleus. 

Some other fission-barrier heights have been also reported for
comparison in Table~\ref{table: fission-barrier heights
  comparison for odd-mass nuclei}. More precisely we consider two sets
of calculations, namely the EFA calculations by Robledo and
collaborators~\cite{Robledo_2009, Iglesia_2009} and the
macroscopic-microscopic calculations by
M\"{o}ller~\cite{Moller_2009}. Three sets of evaluated fission-barrier
heights are alos listed: those fitted to reproduce the neutron-induced
fission cross-sections by Goriely and collaborators~\cite{Goriely_2009},
those coming from the RIPL-3~\cite{Capote_2009} database extracted
from empirical estimates compiled by Maslov \textit{et al.}
\cite{Smirenkin_1993}, and the empirical fission-barrier heights of
Bj\o rnholm and Lynn \cite{Bjornholm_1980} obtained from the lowest-energy solution at
the saddle points irrespective of the nuclear angular-momentum and
parity quantum numbers.

Out of these values, only those obtained from Refs.~\cite{Robledo_2009, Iglesia_2009}
using the Gogny D1S force within the Hartree-Fock-Bogoliubov-EFA framework
are directly comparable with our results. 
In these works, axial symmetry is assumed. 
The resulting fission-barrier heights
are much higher than our calculated values. 
This is consistent
with the rather high fission-barrier heights obtained for the
even-even $^{240}$Pu nucleus in the earlier work of Ref.~\cite{Berger_1984}.

It should be stressed that the rather large differences existing between our results 
and those reported in Refs.~\cite{Robledo_2009, Iglesia_2009} 
cannot be ascribed to the treatment of the
time-reversal symmetry breaking. 
In fact, we have checked that equal-filling approximation (EFA) calculations 
(corresponding to a particle and not quasi-particle blocking though)
affects the total binding energies by a few
hundred keV at most for the parity symmetric case. 
The resuls of calculations for four different configurations
in $^{239}$Pu of $E_A$, $E_{IS}$  and $E_B$ 
are displayed on Table~\ref{table: diff of fission-barrier heights of EFA vs SCB}.
The effect of time-reversal symmetry
breaking terms is found to be approximately constant with deformation. 

\begin{table}[h]
\caption{\label{table: diff of fission-barrier heights of EFA vs SCB}
	Differences (in keV) betwreen the intrinsic fission-barrier heights ($\Delta E_x = \big(E_x\big)_{EFA} - \big(E_x\big)_{SCB}$ with
	$x \equiv A, IS, B$) calculated within
	the EFA and SCB framework for $^{239}$Pu  with the
	SkM* parametrization.}
\begin{ruledtabular}
\begin{tabular}{*{4}c}
$K^{\pi}$&  $\Delta E_A$&    $\Delta E_{IS}$&    $\Delta E_B$  \\
\hline
1/2$^+$&	-70&	-50&	-10	\\
5/2$^+$&	-10&	-20&	0	\\
7/2$^+$&	-10&	-20&	-10	\\
7/2$^-$&	-10&	0&	0	\\
\end{tabular}
\end{ruledtabular}
\end{table}

The comparison with the other sets of data in 
Table~\ref{table: fission-barrier heights comparison for odd-mass nuclei}
is less straightforward. 
As was mentioned by Schunck \textit{et al.} in Ref.~\cite{Schunck_2014}, 
due to an uncertainty in the empirical
fission-barrier heights of about 1 MeV, it is may be illusory
to attempt a reprodution of empirical values within less than such an error bar.
In our case, the fission-barrier heights calculated with the SkM* parametrization
and including the various corrective terms as discussed above, falls easily within this range.

\subsection{Specialization energies \label{Specialization energies}}

Originally (see Refs.~\cite{Wheeler, Newton_1955}), the concept of
specialization energy has been defined as the difference between
fission barrier heights of an odd nucleus with respect to those of
some of its even-even neighbors. Namely one defines, for instance, the
specialization energy for the first (inner) barrier, upon considering
$^{239}$Pu as a $^{238}$Pu core plus one neutron particle, as 
\eq{
\label{eq: specialization energy core-plus-particle}
\Delta E_A^{(p)}( ^{239}\text{Pu} , K^{\pi}) = E_A( ^{239}\text{Pu} ,
K^{\pi}) -  E_A(^{238}\text{Pu} , 0^{+}) \,,
}
and similarly when considering $^{239}$Pu as a $^{240}$Pu core plus
one neutron hole 
\eq{
\Delta E_A^{(h)}( ^{239}\text{Pu} , K^{\pi}) = E_A( ^{239}\text{Pu} ,
K^{\pi}) -  E_A(^{240}\text{Pu} , 0^{+}) \,.
\label{eq: specialization energy core-minus-particle}
}
For configurations at the ground state deformation having a very low
or zero excitation energy, due to the conservation of quantum numbers
preventing to follow the \textit{a priori} lowest energy
configurations at s.p. level crossings, one expects that these
specialization energies should be positive quantities. This is of
course the case for experimentally observed spontaneous fission
processes. But this would not hold whenever one would consider
configurations which correspond to a high enough excitation energy in
the ground state well as we will show in a specific case (see
Table~\ref{table: specialization energy of Pu-239}). 

\begin{figure*}[t]
\includegraphics[angle=-90,keepaspectratio=true,scale=0.6]{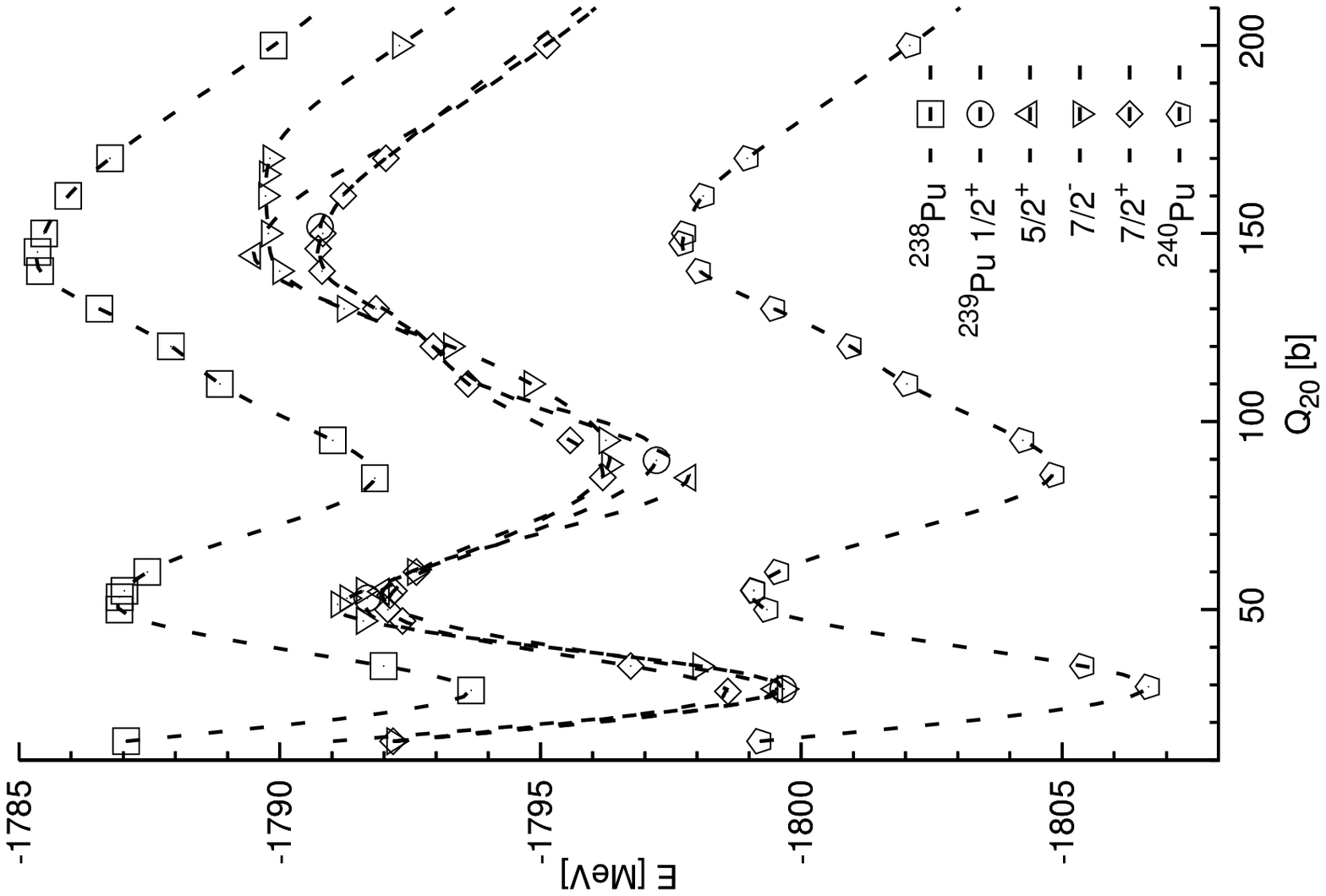}
\includegraphics[angle=-90,keepaspectratio=true,scale=0.6]{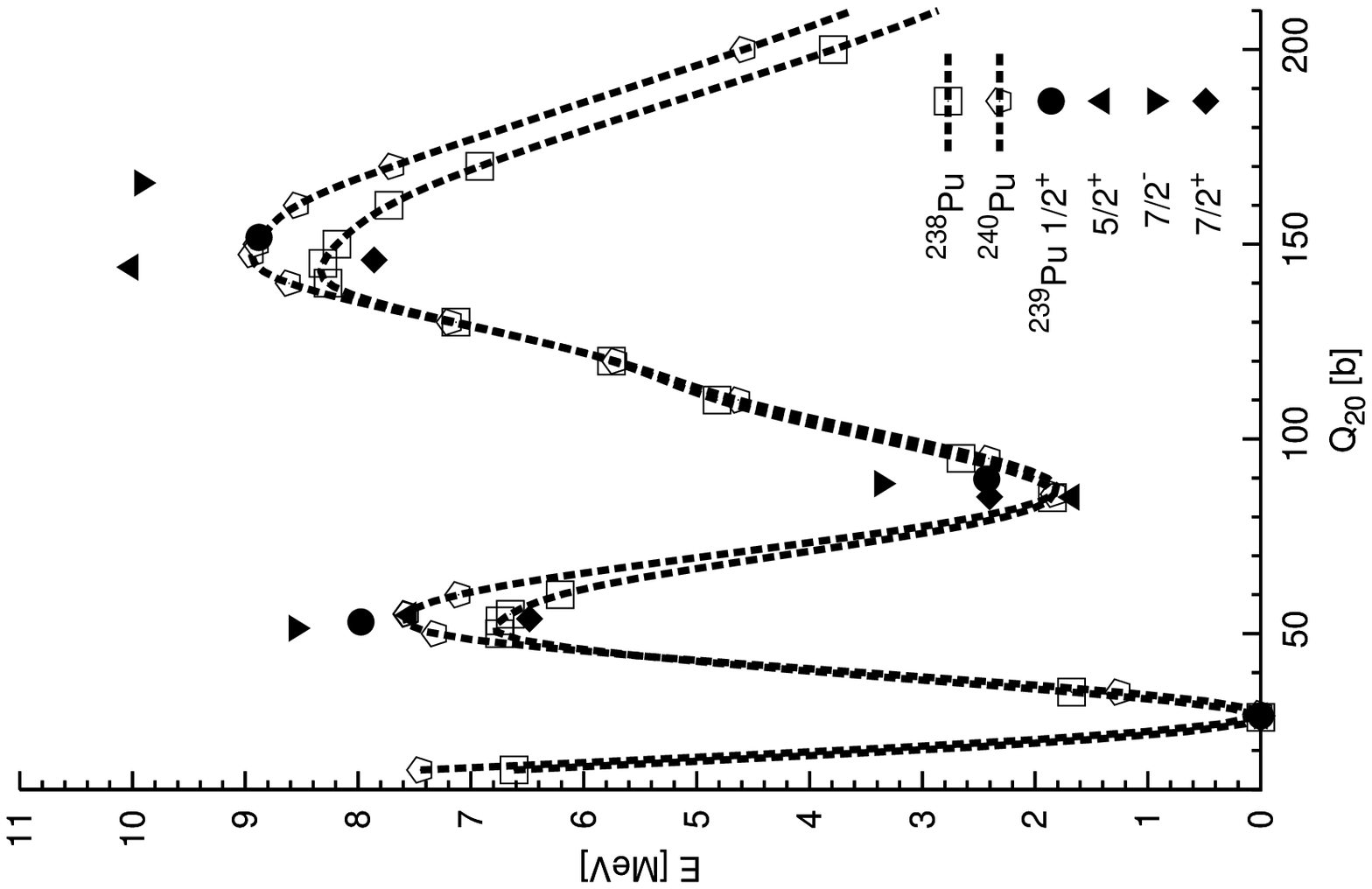}
\caption{\label{fig: specialization energy plutonium} Deformation-energy curves of $^{238,240}$Pu (with $K^{\pi} =
  0^+$) and $^{239}$Pu (with $K^{\pi} = 1/2^+$, $5/2^+$, $7/2^-$ and
  $7/2^+$) as functions of $Q_{20}$ in barns. The Belyaev's moments of
  inertia have been increased by a factor of 2. 
  Left panel: absolute energy scale; right panel: relative
  scale, taking the normal-deformed minimum as the origin of energy
  for all curves.}
\end{figure*}

To illustrate this concept 
Figure~\ref{fig: specialization energy plutonium} and
Table~\ref{table: specialization energy of Pu-239} present the deformation-energy curves and the
fission-barrier heights, respectively, with a conserved parity
symmetry evaluated within the BM unified model for the four blocked
$K^{\pi}$ configurations of $^{239}$Pu with respect to those of the
neighbouring even-even nuclei.
We see that the inner and
outer-barrier heights for some blocked 
configurations---the $7/2^-$ configuration being an excellent
example---are higher than the one of the two
neighboring even-even nuclei as a consequence of
fixing $K^{\pi}$ quantum numbers along the fission path. 
In contrast the $7/2^+$ blocked configuration happens
to yield lower fission-barrier heights as 
compared to the two neighboring even-even nuclei. 
This is so, as above discussed, because the $7/2^+$ configuration is found at
a much higher excitation energy
in the ground-state deformation well~\cite{Koh_2016} but 
with a low excitation energy at the saddle points as
compared to the other blocked configurations. 
This results in negative specialization energies, as shown
in Table~\ref{table: specialization energy of Pu-239}.

\begin{table}[h]
\caption{\label{table: specialization energy of Pu-239} Specialization energies 
  defined here as the average of Eq.~(\ref{eq: specialization energy core-plus-particle}) 
  and (\ref{eq: specialization energy core-minus-particle}) 
  for the four blocked
  configurations of $^{239}$Pu (in MeV). The Belyaev's moments of inertia have been increased
  by a factor of 2.} 
\begin{ruledtabular}
\begin{tabular}{*{8}c}
&  \multicolumn{7}{c}{$K^{\pi}$}  \\
\cline {2-8}
&  1/2$^{+}$&  &  5/2$^{+}$&  &  7/2$^{+}$&  &  7/2$^{-}$  \\
\hline
$\Delta E_{A}$&  0.83&  &  0.39&  &  -0.68&  &  1.41  \\
$\Delta E_{B}$&  0.26&  &  1.38&  &  -0.77&  &  1.31  \\
\end{tabular}
\end{ruledtabular}
\end{table}

By way of conclusion, one can state that the
fission-barrier profiles (heights and widths) are very much dependent
on the $K^{\pi}$ quantum numbers.

\subsection{Effect of neglected time-odd terms 
\label{Effect of neglected time-odd terms}}

In order to probe the effect of the neglected time-odd
  densities we have performed calculations of the total 
binding energy as a function of deformation with parity symmetry  
within the so-called \textit{full time-odd} scheme,
from the normal-deformed ground-state well up to the fission-isomeric well. 
For this study, we have also considered
another commonly used Skyrme parameters set, 
namely the SIII parametrization~\cite{Beiner_1975},
partly because there, the coupling constants $B_{14}$ and
  $B_{18}$ driving the terms involving the spin-current tensor
  density $J_q^{\mu\nu}$ and the Laplacian of the spin density,
  respectively), are exactly zero. In the \textit{full time-odd}
scheme, the $B_{14}$, $B_{15}$, $B_{18}$ and $B_{19}$
coupling constants are not set to zero but allowed to take the 
  values resulting from their expression in terms of the Skyrme
  parameters (see Appendix~\ref{Appendix: Skyrme energy density functional}). 
The contributions to the inner-barrier height $E_A$ and
fission-isomeric energy $E_{\rm IS}$ stemming from the kinetic energy,
the Coulomb energy, the pairing energy as well as the various
coupling-constant terms appearing in the 
Skyrme Hamiltonian density are calculated self-consistently in
the \textit{minimal} and \textit{full time-odd} schemes from our converged solutions.

More specifically we denote by $\Delta E_{B_i}'$ the
  difference between the $B_i$ contribution to the inner-barrier heights
$\Delta E_{B_i}^{(\rm full)}$ and $\Delta E_{B_i}^{(\rm min)}$
in the \textit{full time-odd} and the \textit{minimal time-odd} schemes,
respectively
\eq{
\label{eq_Delta_E'_Bi}
\Delta E_{B_i}'  = \Delta E_{B_i}^{(\rm full)} - \Delta E_{B_1}^{(\rm
  min)} \,.
}
Similarly we denote by $\Delta E_{\rm  kin}^{(\rm full)}$ and $\Delta E_{\rm kin}^{(\rm min)}$ the
kinetic-energy contribution to the inner-barrier height in both time-odd schemes. 
In the same
spirit the abbreviated indices $\rm C$ and $\rm pair$ are used for the
corresponding Coulomb and pairing contributions, respectively.
The sum of the double energy differences
coming from the kinetic, Coulomb, pairing and $B_i$
  contributions with $i$ ranging from 1 to 13
is denoted as $\Delta E_{\min}'$
\eq{
\Delta E_{\min}' = \Delta E_{\rm kin}' + \sum_{i=1}^{13} \Delta E_{B_i}' + \Delta
  E_{\rm pair}' + \Delta E_{\rm C}' \,.
}
The difference of inner-barrier heights in the two \textit{time-odd}
schemes is therefore given by
\eq{
\label{eq_Delta_E'_A}
\Delta E_{A}' = \Delta E_{\min}' + \Delta E_{B_{14}}' + \Delta
  E_{B_{15}}' + \Delta E_{B_{18}}' + \Delta E_{B_{19}}' \,.
}
Similar notations are used for the fission-isomeric energy.

\begin{figure*}[h!]
\includegraphics[angle=-90,keepaspectratio=true,scale=0.75]{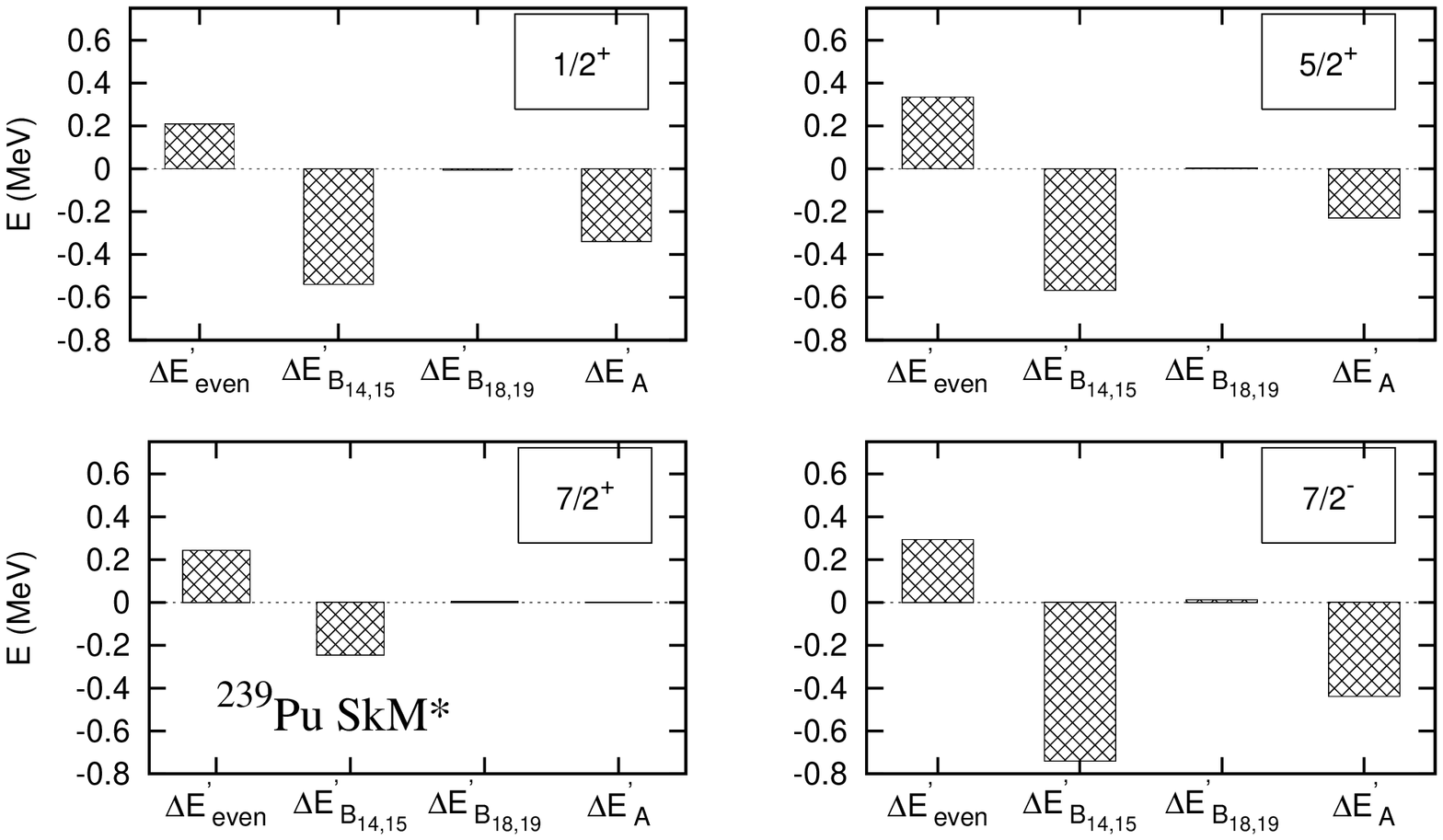}
\includegraphics[angle=-90,keepaspectratio=true,scale=0.75]{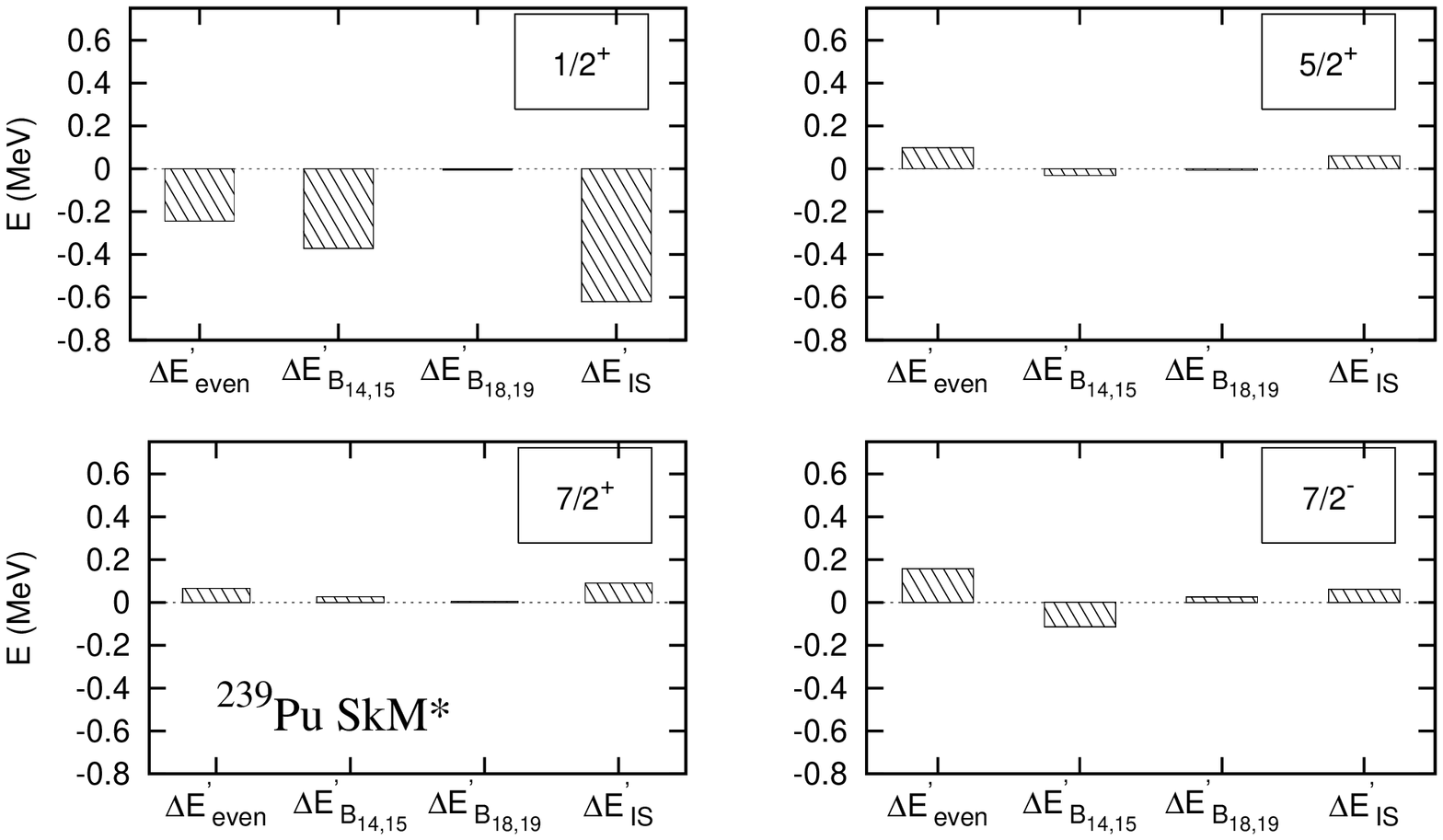}
\caption{\label{fig: Pu239_time_odd_scheme_SkM} Energy differences between various
	contributions 
	(see Eqs.~(\ref{eq_Delta_E'_Bi}) to (\ref{eq_Delta_E'_A}) for definitions) 
	to the inner-barrier height and
	isomeric energy obtained in the default
	\textit{minimal time-odd} scheme and the \textit{full time-odd} scheme
	for several blocked configurations in $^{239}$Pu with
	the SkM* parametrization. The difference in the inner-barrier heights
	$\Delta E_{A}^{'}$ and fission-isomeric energy $\Delta E_{\rm IS}^{'}$
	between the two schemes are also given for each blocked
	configuration.} 
\end{figure*}

\begin{figure*}[h!]
\includegraphics[angle=-90,keepaspectratio=true,scale=0.75]{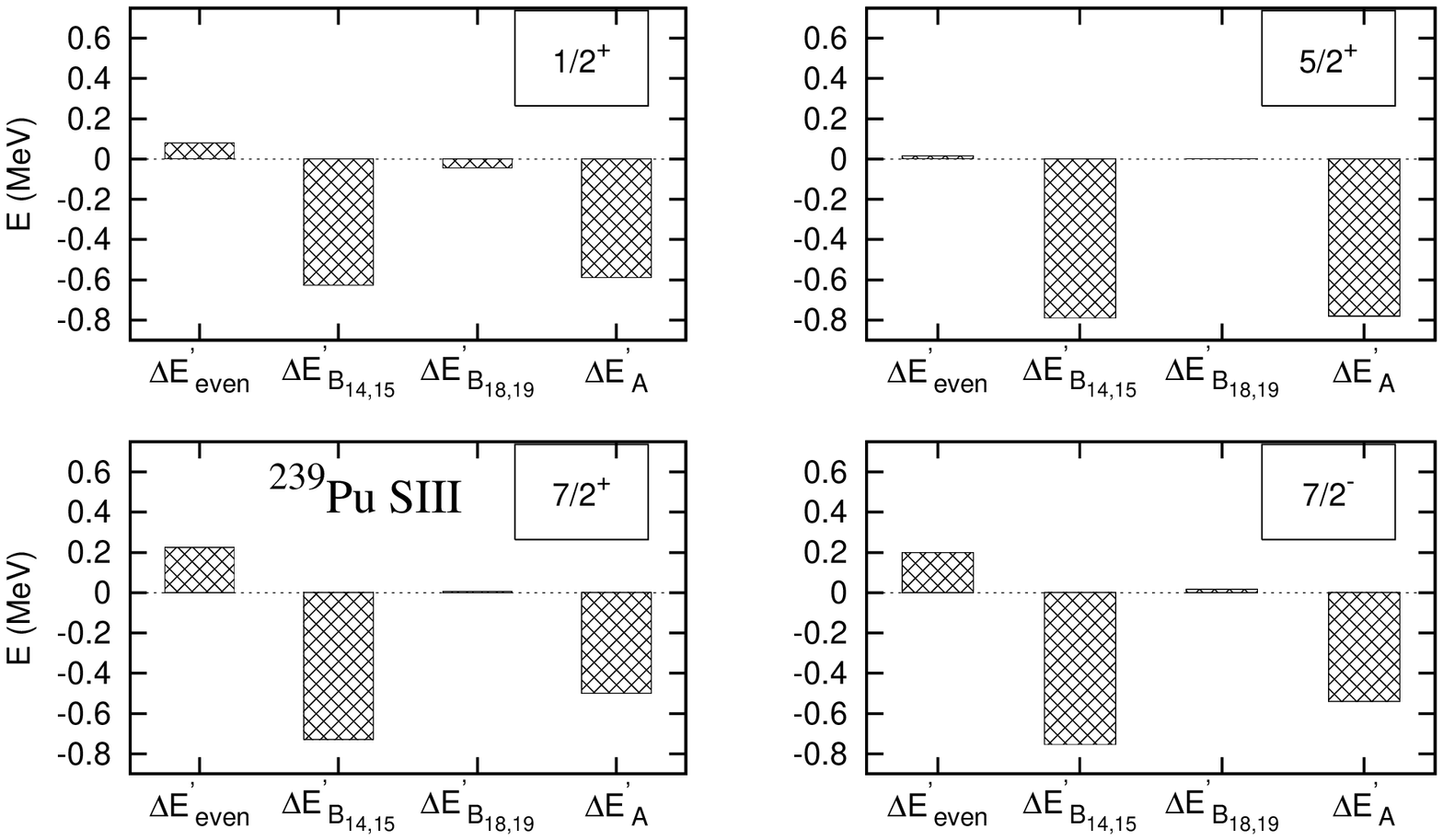}
\includegraphics[angle=-90,keepaspectratio=true,scale=0.75]{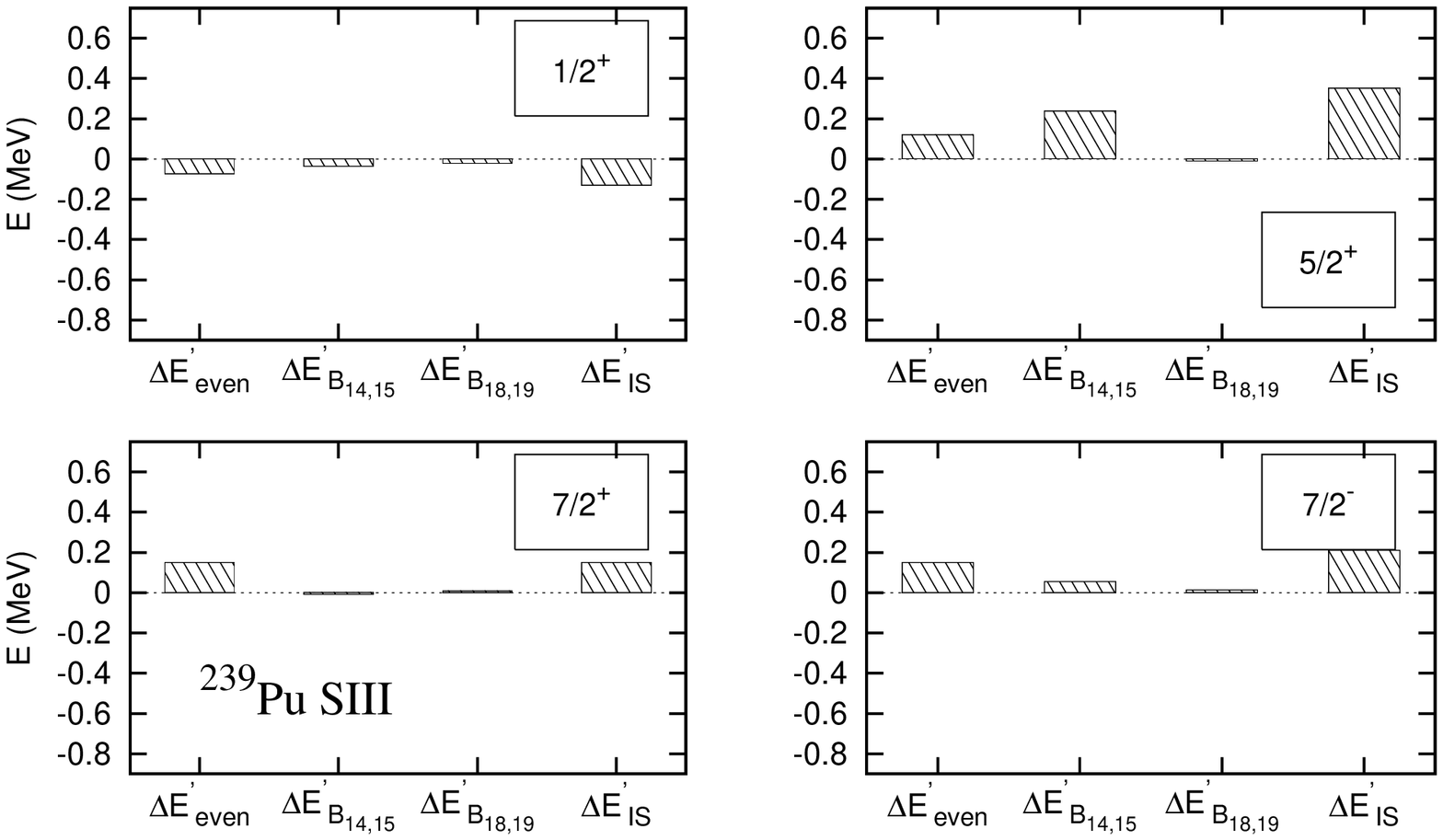}
\caption{\label{fig: Pu239_time_odd_scheme_SIII} Same as Figure \ref{fig: Pu239_time_odd_scheme_SkM}
 for the SIII parametrization.} 
\end{figure*}

In Figures~\ref{fig: Pu239_time_odd_scheme_SkM} 
and \ref{fig: Pu239_time_odd_scheme_SIII}, the various
energy differences defined above, are represented as histograms for the
SkM* and SIII parametrizations, respectively.
We find that the inner-barrier heights, in general, 
decrease when going from a \textit{minimal} to a \textit{full time-odd} 
scheme in all considered blocked configurations. 
This is reflected by the negative values of $\Delta E_A'$. 
The difference in the inner-barrier heights between both
time-odd schemes is overall a competition between the $\Delta
E_{\min}'$ and $\Delta E_{B_{14,15}}'$, while the $\Delta
E_{B_{18,19}}'$ terms have a negligible effect. 
More precisely, the $\Delta E_{B_{14}}'$ term involves the combination
of $\overleftrightarrow{J}^2 - \mathbf{s \cdot T}$
  local densities and is found to be dominated by the
$\overleftrightarrow{J}^2$ component. 
When the $\Delta E_{\min}'$ and
$\Delta E_{B_{14,15}}'$ contributions are of the same magnitude but
with opposite signs, then we do not have a change in the inner-barrier
height, as is the case for the 7/2$^+$ blocked configuration 
with the SkM* parametrization. 

The effect of the time-odd scheme on fission-isomeric energy $E_{\rm IS}$
is less clear-cut. However, we could still observe
that the $B_{18}$ and $B_{19}$ contributions remain  negligible. 
Moreover the time-odd scheme generally has less impact
  on the fission-isomeric energy than on the inner-barrier height. 
A notable exception is found for the 1/2$^+$ configuration.

This study shows that the 
terms proportional to coupling constants which are
not constrained in the original fits
of the Skyrme force can impact the fission-barrier heights in a
non-systematic and non-uniform manner. 
This suggests that one cannot absorb this effect into
an adjustment procedure. 

\section{Spectroscopic properties in the 
fission-isomeric well 
\label{Spectroscopic properties in the fission-isomeric well}}

In this section, we discuss the results obtained in the
fission-isomeric well for the $^{235}$U and $^{239}$Pu nuclei.
We will compare here the results obtained with three Skyrme force parametrizations
(SkM*, SIII and SLy5*).
In the vicinity of the isomeric state, we will make the approximation that the parity mixing is indeed very small, 
such that (with the notation of Subsection II.D)
\eq{\epsilon \sim 1}
and similarly for an odd nucleon state stemming from a positive parity s.p. configuration
\eq{\eta \sim 1  \text{\hspace{1cm}} h^{+} \sim e_{odd} \text{\hspace{1cm}} h^{-} \sim 0  }
while for an odd nucleon state stemming from a negative parity s.p. configuration
\eq{\eta \sim 0  \text{\hspace{1cm}} h^{-} \sim e_{odd}
  \text{\hspace{1cm}} h^{+} \sim 0 \,.  }

As a result for a positive parity nuclear configuration, the projected energy of the fission-isomeric state will be approximated by
\eq{ E(K^{+}) \sim E_{int}(K^{+} ) + \Delta E_{core}^{+}}
while in the negative parity case we will have
\eq{ E(K^{-}) \sim E_{int}(K^{-} ) + \Delta E_{core}^{+}}
where the intrinsic energies $ E_{int}(K^{\pm} ) $ are the energies of our microscopic blocked HF + BCS calculations.

\subsection{Static quadrupole moment
\label{Static quadrupole moments}}

Before discussing relative energy quantities in the fission-isomeric well, 
we assess the quality of deformation properties of our solutions in this well by
calculating the intrinsic quadrupole moments for
some relevant $K^{\pi}$ configurations in the fission-isomeric well. 
The obtained values are listed in 
Table~\ref{table:quadrupole moment in fission isomeric well}.
To the best of our knowledge, experimental values
are available in $^{239}$Pu only~\cite{Habs_1977,Backe_1979}. 
In this nucleus, our values calculated for the $5/2^+$ configuration
with the three considered Skyrme force parametrizations are all
found to agree with experiment within the quoted error bars.  

 \begin{table}[h]
 \caption{\label{table:quadrupole moment in fission isomeric well} Calculated intrinsic 
 quadrupole moments in the isomeric well 
 for the two lowest-energy states 
	in $^{235}$U and the two states corresponding to the 
	experimentally known \cite{Habs_1977,Backe_1979} $K^{\pi}$ configuration in $^{239}$Pu.
	In addition, the values obtained for the $11/2^+$ state in $^{239}$Pu are also reported.}
 \begin{ruledtabular}
 \begin{tabular}{*{9}c}
	Nucleus	   &   $K^{\pi}$&  SkM*&  &   SIII&  &  SLy5*&  &  Exp  \\
	\hline
	\multirow{2}{*}{$^{235}$U}  &	5/2$^+$&   32.9&  &  31.8&  &  33.4&  &   -   \\
		   		    &	11/2$^+$&  32.5&  &  31.8&  &  32.3&  &   -   \\
	\hline
	\multirow{3}{*}{$^{239}$Pu} &   5/2$^+$&   34.1 &  &  33.2 &  &   34.8 &  &  36 $\pm$ 4  \\
				    &   9/2$^-$&   34.1 &  &  33.2 &  &   34.5 &  &  -		\\
				    &   11/2$^+$&  34.5 &  &  33.9 &  &   34.3 &  &  -		\\
 \end{tabular}
 \end{ruledtabular}
 \end{table}

\subsection{Fission-isomeric energy, band heads and rotational bands 
\label{Fission-isomeric energy, band heads and rotational bands}}

\begin{figure*}[h]
\hspace*{-0.75cm}
\includegraphics[angle=-90,keepaspectratio=true,scale=0.75]{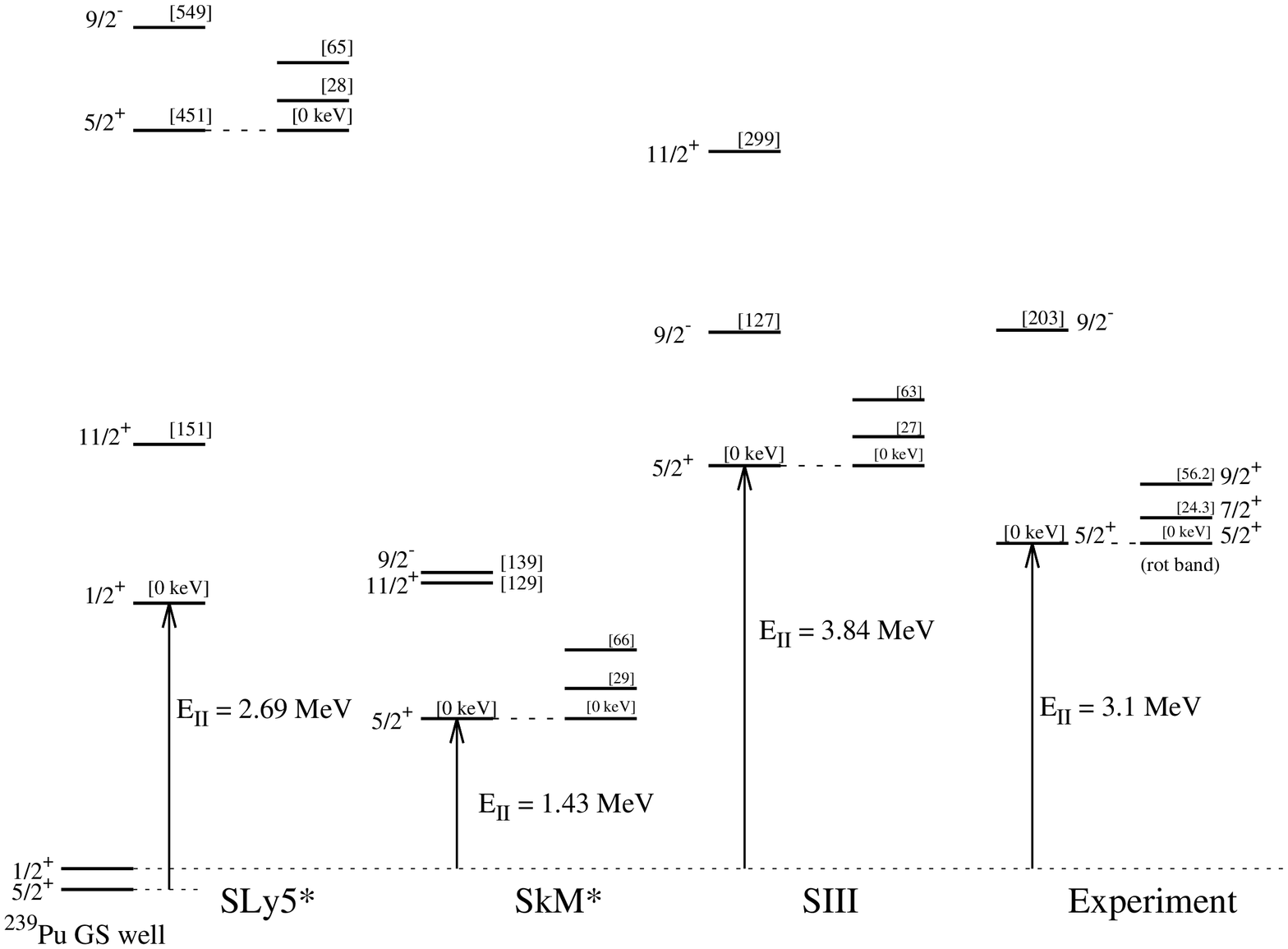}
\caption{\label{fig: Pu239 fission isomeric band head} Band-head
  energy spectra of $^{239}$Pu calculated with the SLy5*, SkM* and
  SIII parametrizations in the isomeric well with the inclusion of the
  rotational correction. The standard Thouless-Valatin correction of 
  Ref. \cite{Libert_Girod_Delaroche_1999} beyond the Belyaev's result has been
  taken into account for the moments of inertia of each
   band. The rotational spectra built upon the lowest-energy
  $5/2^+$ state \textit{(rot band)} are also shown on the second
  column of each Skyrme force. The experimental data are taken from
  Refs. \cite{Browne_2014,Browne_2003}. The fission-isomeric energy
  defined as the energy difference between the lowest-energy solution
  in the ground state and the fission-isomeric well is denoted by
  $E_{\rm II}$.}
\end{figure*}

\begin{figure*}[h]
\hspace*{-1cm}
\includegraphics[angle=-90,keepaspectratio=true,scale=0.75]{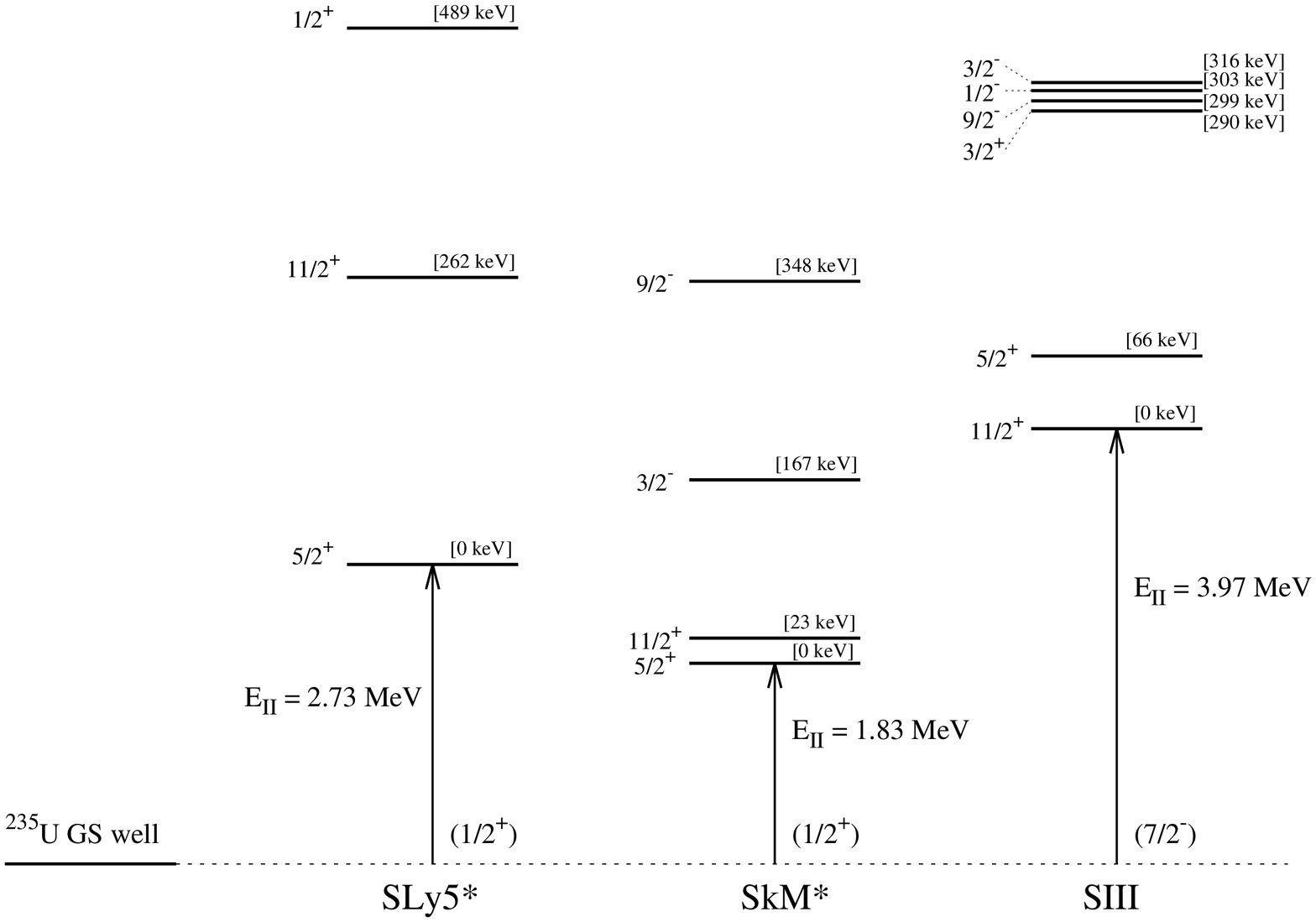}
\caption{\label{fig: U235 fission isomeric band head} Same as Figure~\ref{fig: Pu239 fission isomeric band head} 
	 for $^{235}$U.}
\end{figure*}

Above the lowest-energy solution at the fission-isomeric well
there are several band-head states within 1 MeV. 
This has been displayed in 
Figures~\ref{fig: Pu239 fission isomeric band head} 
and \ref{fig: U235 fission isomeric band head}
for the $^{239}$Pu and $^{235}$U nuclei, respectively.
These results have been obtained with the inclusion of rotational energy
with the approximate Thouless-Valatin
corrective term in the moment of inertia (assuming a 32 \% increase above the IB value). 

Let us first discuss the energy spectra for the $^{239}$Pu nucleus
for which a comparison with the experimental data of 
Refs.~\cite{Browne_2014,Browne_2003} is possible. 
As shown in Fig.~\ref{fig: Pu239 fission isomeric band head},
the experimental ground-state quantum numbers 
in the normal-deformed well are
$1/2^+$ while in the fission-isomeric well they are $5/2^+$.
Our calculated results with the SkM* and the SIII parametrizations reproduce
these data. 

On the contrary, the calculations with the SLy5* parameter set, fail
to do it as they yield a $5/2^+$ ground-state in the normal-deformed
well located 160 keV below the 1/2$^+$ state and a $1/2^+$
lowest-energy state in the fission-isomeric well. 
Moreover, the $K^{\pi} = 9/2^-$ state calculated with SLy5* appears at
a too high excitation energy of more than 500~keV as compared to the
experimental value of about 200~keV. 

In contrast the excitation energy of this $9/2^-$ state is found in
much better agreement with data for SkM* and SIII (139 keV and 127
keV, respectively). The agreement with the data of these values is
expected to be favorably improved when including the effect of
Coriolis coupling, as suggested from the work of
Ref.~\cite{Libert_1980}. In addition, a $11/2^+$ excited state is
predicted at 151 keV, 129 keV and 299 keV with the SLy5*, the SkM* and
the SIII parametrizations, respectively. This state was also predicted
(at a 44~keV excitation energy) in the Hartree--Fock--Bogoliubov
calculations with the Gogny force by Iglesia and collaborators~\cite{Iglesia_2009}.  

The rotational band built on the $5/2^+$ band-head state can also be
compared with experimental data: the calculated energies for the first
two excited states are found to be rather similar within the three
considered Skyrme parametrizations in use, and to compare very well
with data.

Let us now move the discussion to the results for the $^{235}$U nucleus
displayed in Fig.~\ref{fig: U235 fission isomeric band head}.
To the best of our knowledge, there are
no experimental data available for comparison with our calculated
values in the superdeformed well of this nucleus. There are, however,
some calculations performed with the Gogny force in the work of
Ref.~\cite{Robledo_2009} which predict a $5/2^+$ ground state with a
first 11/2$^+$ excited state at 120 keV in the fission-isomeric
well. The same level sequence is also obtained in our calculations
with the SkM* and the SLy5* Skyrme parametrizations, although the
11/2$^+$ state is located at a much higher energy in the latter
parametrization. The calculations with SIII yields the opposite level
sequence, with a 5/2$^+$ state 66 keV above the 11/2$^+$ ground-state.

\subsection{Fission-isomeric energies}

Let us discuss now the fission-isomeric energy $E_{\rm II}$. 
Table~\ref{table: effect of moment of inertia on fission isomeric energy} 
displays the fission isomeric energies $E_{\rm II}$  defined as the
difference between the energies of the solutions lowest in energy in
both the ground state and fission-isomeric wells (irrespective of
their $K^{\pi}$ quantum numbers), namely with an obvious notation 
\eq{
E_{\rm II} = E_{0}^{\rm IS} - E_{0}^{\rm GS} \,.
}
As seen on Table~\ref{table: effect of moment of inertia on fission isomeric energy} 
(see also Figs.~\ref{fig: Pu239 fission isomeric band head} 
and~\ref{fig: U235 fission isomeric band head}) when using the
standard Thouless-Valatin correction of $32 \%$ over the IB estimate,
the Skyrme SIII interaction yields values of $E_{\rm II}$  which are
much too high. This is not very surprising in view of the well-known
defect of its surface tension property. On the contrary, the too low
value obtained with the SkM$^{*}$ interaction which provides very good
Liquid Drop Model barrier heights must be explained by some inadequate
account of relevant shell effect energies. The last interaction
(SLy5$^{*}$) provides reasonable $E_{\rm II}$ values (yet slightly too
weak).

Now, as discussed before, rotational energy corrections calculated
using the Belyaev moment of inertia were found to be too large,
resulting in an underestimation of the fission-barrier heights. This
is partly due to the resulting overestimation of the rotational
correction. As a rough cure for this, one may increase the IB moments
of inertia by a factor of 2. The resulting $E_{\rm II}$ values are listed in
Table~\ref{table: effect of moment of inertia on fission isomeric energy}. 
\begin{table*}
\caption{Fission-isomeric energy $E_{\rm II}$ for three different
prescriptions for the moment of inertia. The $K^{\pi}$ quantum numbers of the
ground-state solution in the fission-isomeric well are those displayed in 
Figs.~\ref{fig: Pu239 fission isomeric band head} and 
\ref{fig: U235 fission isomeric band head},
except for $^{235}$U with the SkM* parametrization and when increasing the Belyaev's result by a factor of 2
(column labeled IB+100\%),
for which the $K^{\pi} = 11/2^+$ blocked configuration
has been considered.
}
\begin{ruledtabular}
\begin{tabular}{*{14}c}
\multirow{2}{*}{Nucleus}&  \multicolumn{3}{c}{SLy5*}&  & \multicolumn{3}{c}{SkM*}&  &  
	\multicolumn{3}{c}{SIII}&  &  \multirow{2}{*}{Exp} \\
\cline{2-4} \cline{6-8} \cline{10-12}
	&	IB&  IB+32\%&  IB+100\%&  &  IB&  IB+32\%&  IB+100\%&  &  IB&  IB+32\%&  IB+100\%&  \\
\hline
$^{235}$U&  2.36&  2.73&  3.11&  &  1.46&  1.83&  2.20&  &  3.62&  3.97&  4.35&  &   -  \\
$^{239}$Pu& 2.30&  2.69&  3.10&  &  1.08&  1.43&  1.80&  &  3.42&  3.84&  4.30&  &  3.1 \\
\end{tabular}
\label{table: effect of moment of inertia on fission isomeric energy}
\end{ruledtabular}
\end{table*}
It has been checked that the band-head energy spectra in
the fission-isomeric well are then only affected by
some tens of keV from the values shown in 
Figs.~\ref{fig: Pu239 fission isomeric band head} and 
\ref{fig: U235 fission isomeric band head}. 
The $K^{\pi}$ quantum numbers of the
lowest-energy solutions in all cases remain unchanged 
except for $^{235}$U with the SkM* interaction. In this case, we have
a change in the level ordering of the ground and first excited
states, where the quoted value of $E_{\rm II} = 2.20$~MeV involves 
the $K^{\pi} = 11/2^+$ blocked
configuration  in the fission-isomeric well.

\section{Concluding remarks\label{Conclusion}}

From the above calculations of fission barriers in odd-mass
nuclei within a self-consistent blocking approach we can draw the
following conclusions.

First, barrier heights and fission isomeric energies depend on the time-odd
scheme in a non-systematic way. Indeed they are found to vary with the
nucleus and with the quantum numbers in a given nucleus between zero
and almost 0.8 MeV in the studied nuclei. This effect cannot be
absorbed in the adjustment of the Skyrme parameters. In particular the
calculated specialization energies strongly vary with the $K$ and
$\pi$ quantum numbers and can be negative when the blocked
configuration lies rather high in energy in the ground-state well and
rather low at the saddle point.

Moreover, the equal-filling approximation, defined in our work as an
equal occupation of the block single-particle state and its
time-reversed state as opposed to the definition of
Ref.~\cite{Schunck_2010} based on one-quasi-particle states, is found to
have no significant effect on deformation and is a fairly good
approximation to calculate relative energies, such as the
fission-barrier heights and fission-isomeric energies.

Regarding spectroscopic properties in the ground-state and
fission-isomeric wells, we have found overall a fair agreement with
available data. This gives us some confidence in the deformation
properties of the fissioning nuclei, especially in the barrier
profiles as functions of the $K^{\pi}$ quantum numbers. 

In this context, we recall that we have imposed axial symmetry throughout the whole potential energy curve so that the $K$ quantum number remains meaningful. As already discussed, this may be deemed as a reasonable assumption in view of dynamical calculations performed for $^{240}$Pu and heavier nuclei, showing that the least-action path
is closer to an axial path than the triaxial static one around the top
of the inner and outer barriers
\cite{Gherghescu99,Sadhukhan14,Zhao16}. Moreover, as far as class I
states are concerned, it has been established from gamma decay of
even-even and odd-odd rare earth nuclei formed by neutron capture,
that the K-quantum number is reasonably conserved even at energies
resulting from neutron capture in the thermal and resonance energy
domains (see, e.g., \cite{OSLO}).

Regardless of the validity of the axial symmetry assumption, our
calculations of fission barriers with fixed $K$ values allow to expect
that the penetrabilities of inner and outer fission barriers will
strongly vary with the blocked configurations, resulting in a
widespread distribution of fission-transmission coefficients as a
function of $K$ and $\pi$ for a fixed $J$ quantum number. This can a
priori impact the fission cross section computed in the optical model
for fission with the full $K$ mixing approximation (see, for instance,
\cite{Sin06,Sin08}).

\begin{figure*}
\hspace*{-0.5cm}
\includegraphics[angle=-90,keepaspectratio=true,scale=0.7]{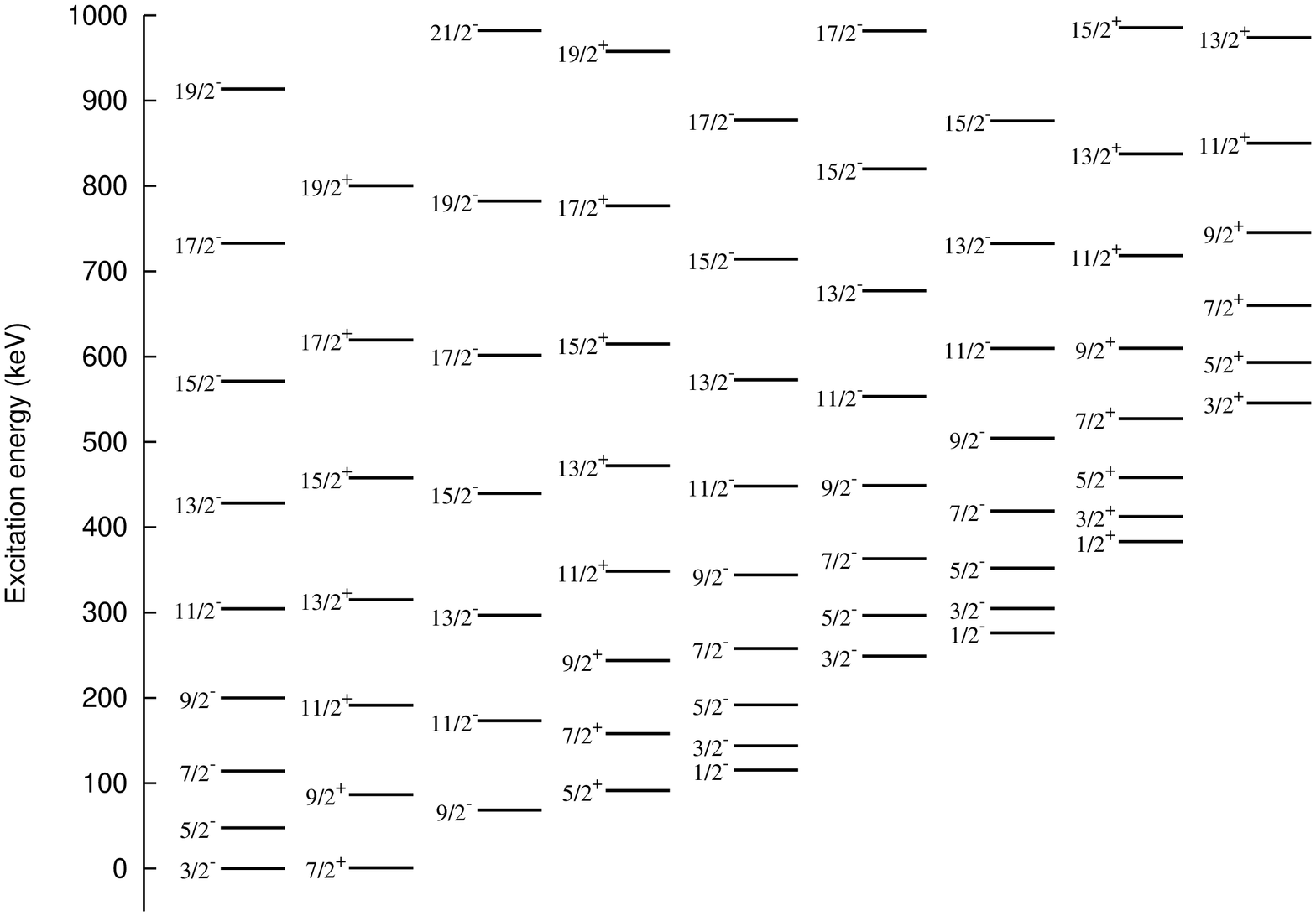}
\caption{Example of calculated transition states at the top of the inner barrier of $^{239}$Pu.
\label{fig_transition_states}}
\end{figure*}

As a matter of fact fission cross section calculations require in 
principle the knowledge of penetrabilities for each discrete
transition state, that is the barrier profile and inertia parameters
for each discrete state at barrier tops. In
Fig.~\ref{fig_transition_states} we show such transition states as
rotational bands built on various low-lying blocked
configurations. They are calculated in the above discussed
Bohr--Mottelson approach using Skyrme-HFBCS intrinsic solutions with 
self-consistent blocking. 
This kind of results can provide microscopic input to the discrete
contribution to the fission transmission coefficients, along the lines
of Ref.~\cite{Goriely_2009}. Note that, in this work, odd-mass nuclei
were not considered in a time-reversal symmetry breaking approach and
that the inertia parameters were calculated within a hydrodynamical
model. A natural extension, requiring very long computing times, is to
compute these parameters from a microscopic model as in the non-perturbative ATDHFB  approach~\cite{Yuldashbaeva_1999},
consistently with the barrier profiles for each blocked configuration.

Finally, it is to be noted that in such dynamical calculations, and
even in static calculations, the phenomenological quality of the
pairing interaction is of paramount importance. In our case, its
intensities have been determined by a fit based on explicit
calculations of odd-even mass differences in the actinide
region. However, such approaches suffer a priori from the deficiencies
inherent to a non-conserving particle-number theoretical framework,
particularly so if strong pairing fluctuations are to be
considered. To cure for that in an explicit and manageable fashion, we
intend to perform similar calculations as those presented here, using
the so called Highly Truncated Diagonalization Approach of
Ref.~\cite{Pillet_2002}. 

\appendix

\section{Skyrme energy density functional \label{Appendix: Skyrme energy density functional}}

As well known, when using an effective internucleon interaction of the Skyrme type,
the total energy of a normalized Slater determinant $\ket{\Psi_{HF}}$
can be written as an integral of a Hamiltonian density, $\mathcal{H}$, such that:
\begin{widetext}
\eq{
E = \elmx{\Psi_{HF}}{\hat{H}}{\Psi_{HF}}  
  = \int \mathcal{H} (\mathbf{r}) \; d\mathbf{r} 
  = \int \Big( \mathcal{H}_{kin}(\mathbf{r}) + \mathcal{H}_c(\mathbf{r})
  + \mathcal{H}_{DD}(\mathbf{r}) + \mathcal{H}_{s.o}(\mathbf{r}) + \mathcal{H}_{Coul}(\mathbf{r}) \Big) d\mathbf{r} 
\label{eq: Skyrme total energy}
}
\end{widetext}
where the various Hamiltonian densities $\mathcal{H}_{kin}$,
$\mathcal{H}_c$, $\mathcal{H}_{DD}$, $\mathcal{H}_{s.o}$ and
$\mathcal{H}_{Coul}(\mathbf{r})$ are given
\cite{Engel_1975,Bonche_1987} by (see Table~\ref{coupling constants}
for the definition of the coefficients $B_i$ as function of the usual
$t_i$, $x_i$ and $W_0$ parameters of the Skyrme interaction in use)
\begin{widetext}
\begin{flalign}
    \mathcal{H}_{kin}(\mathbf{r}) =& \; \Big( 1-\frac{1}{A} \Big) \frac{{\hbar}^{2}}{2m} \tau \\
    \mathcal{H}_c(\mathbf{r}) =& \; B_1 \rho^2 + B_{10} {\bf{s}}^2 + B_3 (\rho \tau - {\bf{j}}^2)
    + B_{14}({\overleftrightarrow{J}}^2 - {\bf{s}} \cdot {\bf{T}}) + B_5 \rho {\bf{\triangle}} \rho 
    + B_{18} {\bf{s}} \cdot {\bf{\triangle s}}  \notag \\
    &+ \sum_q \{ B_2 \rho_q^2 + B_{11} {\bf{s}}_q^2 
    + B_4 (\rho_q \tau_q - {\bf{j}}_q^2)
    + B_{15} \big( {\overleftrightarrow{J}_q}^2 - {\bf{s}}_q \cdot {\bf{T}}_q \big) \}
    + B_6 \rho_q {\bf{\triangle}} \rho_q + B_{19}{\bf{s}}_q \cdot {\bf{\triangle s}}_q \\
    \mathcal{H}_{DD}(\mathbf{r}) =& \; \rho^{\alpha} \Big[ B_7
    {\rho}^2 + B_{12} {\bf{s}}^2 + \sum_q (B_8 \rho_q^2 + B_{13} {\bf{s}}_q^2) \Big] \\
    \mathcal{H}_{s.o}(\mathbf{r}) =& \; B_9 \Big[ \rho {\bf{\nabla \cdot J}} + {\bf{j \cdot \nabla \times s}} 
    + \sum_q \big( \rho_q {\bf{\nabla}} \cdot {\bf{J}}_q + {\bf{j}}_q
    \cdot {\bf{\nabla}} \times {\bf{s}}_q \big) \Big] \\ 
    \mathcal{H}_{Coul}(\mathbf{r}) \approx& \; \frac{1}{2} \rho_p(\mathbf{r})V_{CD}(\mathbf{r}) 
    - \frac{3}{4}e^2 {(\frac{3}{\pi})}^{\frac{1}{3}} \rho_p^{\frac{4}{3}}(\mathbf{r}) \label{eq:Coulomb density}
\end{flalign}
\end{widetext}

\begin{table*}
\caption{\label{coupling constants} Definition of the coupling constants $B_i$ entering the expression of Hamiltonian densities, in terms of usual Skyrme force parameters.
}
\begin{ruledtabular}
\renewcommand{\arraystretch}{2.5}
\begin{tabular}{*3l}
$\displaystyle B_1 = \frac{t_0}{2}\big(1+\frac{x_0}{2}\big)$ &
        $\displaystyle B_2 = -\frac{t_0}{2}\big(\frac{1}{2}+x_0\big)$ &
		$\displaystyle B_3 = \frac{1}{4}\big[t_1\big(1+\frac{x_1}{2}\big) + t_2 \big(1+\frac{x_2}{2}\big)\big]$ \\
$\displaystyle B_4 = -\frac{1}{4} \big[t_1 \big(\frac{1}{2}+x_1 \big) - t_2 \big( \frac{1}{2}+x_2\big) \big]$ &
	$\displaystyle B_5 = -\frac{1}{16}\big[3t_1\big(1+\frac{x_1}{2}\big)-t_2\big(1+\frac{x_2}{2}\big)\big]$&
	       $\displaystyle B_6 = \frac{1}{16}\big[3t_1\big(\frac{1}{2}+x_1\big)+ t_2\big(\frac{1}{2}+x_2\big)\big]$ \\
$\displaystyle B_7 = \frac{t_3}{12}\big(1+\frac{x_3}{2}\big)$ &
        $\displaystyle B_8 = - \frac{t_3}{12}\big(\frac{1}{2}+x_3\big)$&
		$\displaystyle B_9 = - \frac{W_0}{2}$ \\
$\displaystyle B_{10} = \frac{1}{4}t_0x_0$ &
	$\displaystyle B_{11} = - \frac{1}{4} t_0$&
               $\displaystyle B_{12}=\frac{1}{24}t_3x_3$ \\
$\displaystyle B_{13} = - \frac{t_3}{24}$ &
        $\displaystyle B_{14} = -\frac{1}{8}\big(t_1x_1+t_2x_2\big)$ &
		$\displaystyle B_{15} = \frac{1}{8}\big(t_1 - t_2\big)$ \\
$\displaystyle B_{18} = -\frac{1}{32}\big(3t_1x_1-t_2x_2\big)$ &
	$\displaystyle B_{19} = \frac{1}{32}\big(3t_1+t_2\big)$ &
\end{tabular}
\end{ruledtabular}
\end{table*}

The factor $(1-\frac{1}{A})$ appearing in the kinetic energy density 
is a corrective term introduced to approximately eliminate the center-of-mass motion spuriously introduced by the 
breaking of the translational invariance inherent to the mean-field approach.
Such an approach has been noted to overestimate 
the contribution from the center-of-mass
correction \cite{Bender_2003}.
Nevertheless, the approximate treatment of the correction term 
is consistent with the manner in which the adopted Skyrme 
parametrizations were fitted.
For a study on the various approximations of the center-of-mass correction in the mean-field approach
and also its effects on nuclear properties as  
deformation energy surface we refer to
Ref.~\cite{Bender_2000}.

The direct part of the Coulomb mean field $V_{CD}$ is readily calculated from the proton density 
(see for the numerical method in use, e.g. Refs. \cite{Vautherin_1973, Samsoen_1999, Ludovic_thesis_2003}).
The exchange part given by the second term of equation (\ref{eq:Coulomb density})
has been approximated here as usually done,  with a Slater approximation \cite{Slater_1951}.
The effect of using such an approximation as opposed to performing
rather time-consuming exact Coulomb exchange calculations have been previously
investigated (see Ref. \cite{Titin_1974, Skalski_2001, Bloas_2011}).
It has been found that the appropriateness of the Slater approximation is 
directly related with the proton single-particle level density near the Fermi level,
being less good for a spherical (close shell) nucleus as compared to a well deformed nucleus.
Consequently, the lowering of the total energy is lesser at the top of the barrier 
due to a higher single-particle level density when the Slater approximation is more appropriate
as compared to the ground-state solution.
This translates into an underestimation of the fission-barrier heights 
when calculations of the Coulomb exchange term are performed using the Slater approximation.

All the above Hamiltonian densities  are time-even functionals of the local densities that are
further categorized into time-even and time-odd densities 
with respect to the action of the time-reversal operator.
The time-even densities are the particle density $\rho(\mathbf{r})$, the kinetic energy density $\tau(\mathbf{r})$
and the spin-current density $J_{\mu \nu}(\mathbf{r})$ 
whose explicit definition can be found in Refs. 
\cite{Engel_1975,Bonche_1987}.

For each of the time-even densities, there exists a time-odd counterpart, namely
the spin density ${\bf{s}}(\mathbf{r})$, the spin kinetic density, ${\bf{T}}_{\mu}(\mathbf{r})$
and the current density ${\bf{j}}(\mathbf{r})$ (see Refs. \cite{Engel_1975,Bonche_1987}).

The Hartree-Fock equations obtained by varying the total energy given in equation (\ref{eq: Skyrme total energy})
with respect to the single-particle wavefunctions $\phi_k$ yield the
following one-body Hamiltonian $\hat{h}_{HF}$~\cite{Engel_1975,Bonche_1987}
\begin{widetext}
\begin{flalign}
    \langle \mathbf{r} |\hat{h}_{HF}^{(q)}|\phi_k\rangle
    =& \; - {\bf{\nabla}}\cdot \Big( \frac{{\hbar}^2}{2m_q^*(\mathbf{r})}{\bf{\nabla}}[\phi_k](\mathbf{r})\Big)
    + \Big(U_q(\mathbf{r})+\delta_{qp}U_{Coul}(\mathbf{r})\Big)[\phi_k](\mathbf{r}) 
    + i{\bf{W}}_q(\mathbf{r})\cdot \Big({\bf{\sigma}} \times {\bf{\nabla}}[\phi_k](\mathbf{r})\Big) \notag\\
    &- i \sum_{\mu,\nu} \Big\{ \Big(W_{q,\mu \nu}^{(J)}(\mathbf{r})\sigma_{\nu} {\bf{\nabla}}_{\mu}[\phi_k](\mathbf{r})\Big) 
    +{\bf{\nabla}}_{\mu}\Big(W_{q,\mu \nu}^{(J)}(\mathbf{r}) \sigma_{\nu}[\phi_k](\mathbf{r}) \Big) \Big\}
    -\frac{i}{2}\Big\{ {\bf{A}}_q(\mathbf{r}) \cdot {\bf{\nabla}}[\phi_k](\mathbf{r}) \notag\\
    &+ {\bf{\nabla}} \cdot \Big( {\bf{A}}_q[\phi_k](\mathbf{r}) \Big)\Big\} 
    + {\bf{S}}_q(\mathbf{r})\cdot{\bf{\sigma}}[\phi_k](\mathbf{r}) 
    - {\bf{\nabla}} \cdot \Big( \big( {\bf{C}}_q(\mathbf{r}) \cdot {\bf{\sigma}} \big){\bf{\nabla}}[\phi_k](\mathbf{r}) \Big)
\label{eq: HF equation}
\end{flalign}
\end{widetext}

The fields $m^*$, $U_q$, $U_{Coul}$, ${\bf{W}}_q$ and $W_{q, \mu \nu}^{(J)}$ 
where the notation $q$ labels the nuclear charge state, are time
even-operators wheras the fields ${\bf{S}}_q$, ${\bf{A}}_q$ and
${\bf{C}}_q$ are time-odd operators. They are given as
follows~\cite{Engel_1975,Bonche_1987} in terms of the various
densities by 
\begin{widetext}
\begin{flalign}
   \frac{{\hbar}^2}{2m_q^*} =& \; \frac{{\hbar}^2}{2m_q} + B_3\rho + B_4\rho_q \\
   \notag\\
   U_q =& \; 2 \big(B_1 \rho + B_2 \rho_n \big) + B_3\tau + B_4 \tau_q
   + 2 \big(B_5\Delta\rho + B_6\Delta\rho_q \big) + (2+ \alpha)B_7\rho^{1+\alpha} \notag \\
   &+ B_8 \big(\alpha \rho^{(\alpha-1)} \big(\rho_n^2+\rho_p^2\big)+2\rho^{\alpha}\rho_q \big)
   + B_9 \big( {\bf{\nabla \cdot J}} + {\bf{\nabla \cdot J_q}}\big)
   + \alpha \rho^{\alpha-1} \big(B_{12} {\bf{s}}^2 + B_{13} \big({\bf{s}}_n+{\bf{s}}_p^2\big) \big) \\
   \notag\\
   U_{Coul} =& \; V_{dir} - e^2 \; \big( \frac{3}{\pi} \rho_p  \big)^{1/3} \\
   \notag\\
   {\bf{W}}_q =& \; - B_9 \big( {\bf{\nabla}}\rho + {\bf{\nabla}}\rho_q \big) \\
   \notag\\
   W_{q,\mu \nu} =& \; B_{14} J_{\mu \nu} + B_{15} J_{q, \mu \nu}  \\
   \notag\\
   {\bf{S}}_q =& \; 2 \big(B_{10} + B_{12} \rho^{\alpha}\big){\bf{s}} + 2 \big(B_{11}+B_{13}\rho^{\alpha}\big){\bf{s}}_q
              -B_9 {\bf{\nabla}} \times \big({\bf{j}}+{\bf{j}}_q\big) - B_{14}{\bf{T}}
              -B_{15}{\bf{T}}_q + 2 \big(B_{18}{\Delta s} + B_{19}{\Delta s}_q\big) \\
   \notag\\
   {\bf{A}}_q =& \; -2 \big(B_3\;{\bf{j}}+B_4\;{\bf{j}}_q\big) + B_9 {\bf{\nabla}} \times \big( {\bf{s}}+{\bf{s}}_q \big) \\
   \notag\\
   {\bf{C}}_q =& \; - \Big( B_{14}\;{\bf{s}} + B_{15}\;{\bf{s}}_q \Big)
\end{flalign}
\end{widetext}

\section{Effect of basis size on fission-barrier heights 
\label{Appendix: Effect of basis size on fission-barrier heights}}

The single-particle states of the canonical basis are expanded 
on deformed harmonic oscillator basis states
truncated according to the deformation-dependent truncation scheme of Ref. \cite{Flocard_1973}.
>From the oscillator frequencies $\omega_{\bot} , \omega_z$ one defines a spherical frequency 
$\omega_0$ by $\omega_0^3 = \omega_z \omega_{\bot}^2 $. The corresponding 
basis parameters $b = \sqrt{\frac{m \omega_0}{\hbar}}$ and $q =
\frac{\omega_{\bot}}{\omega_z}$ are optimized to yield the minimal
energy given a basis size, $N_0$. For computational time reasons,
the  calculations are performed with a basis size defined by $N_0
= 14$ corresponding to 15 major shells in the spherical case.
The \textit{b} and \textit{q} parameters for the calculations
involving the SIII and SkM* interactions in odd-mass nuclei are
deduced as an average of the optimized basis of its neighbouring
even-even isotopes, at each deformation points. It was furthermore
checked that the optimal parameter values obtained for the SkM*
interaction were applicable for the SLy5* parameters sets up to a very
good approximation (of the order of tens of keV).

In this Appendix we assess the basis size effect on fission-barrier
heights using the notation of Subsection II.E. In practice we have
performed such a study for the $^{239}$Pu and $^{240}$Pu nuclei,
assuming axial and parity symmetry along the whole fission path.
Calculations were performed with the SkM* interaction for three basis
sizes ($N_0$ = 14, 16, 18). It would be a priori desirable to optimize
the \textit{b} and \textit{q} parameters for each basis size. However,
the work of Ref.~\cite{Bonneau_2004} comparing solutions which has
been optimized in their respective basis size, has shown that the
impact of the optimization process on the barrier heights is rather
small in determining the considered basis size effect. Thus, the same 
b and q parameters obtained in the optimization process in $ N_0 = 14$
have been used for other $N_0$ values. The locations of the saddle
points as well as the ground states and second minima in the
deformation energy surface were obtained by using the modified
Broyden's method~\cite{Baran_2008}. 

The fission-barrier heights obtained for the various basis sizes
are shown in Table~\ref{table: barrier heights in basis size}. The
truncation effect are shown to increase with deformation. As a result
we have crudely estimated for all considered nuclei (even-even or odd)
that the calculations performed with a basis defined by  $N_0 = 14$
overestimate on the average, the inner-barrier height by about 300 keV
and the isomeric energy as well as the outer-barrier height by about
500~keV. 

\begin{table}
\caption{\label{table: barrier heights in basis size}  
	Inner-barrier $E_A$, fission-isomeric energy ($E_{II}$ for even-even nucleus
	and $E_{IS}$ for odd-mass nucleus) and
	outer-barrier $E_B$ heights of $^{239}$Pu and $^{240}$Pu assuming axial and 
	intrinsic parity symmetries
	for different harmonic oscillator basis sizes calculated with the SkM* interaction. 
	All energies are given in MeV.}
\begin{ruledtabular}
\begin{tabular}{*{8}c}
Nucleus&  $N_0$& &	$E_A$& &	$E_{IS}$ /$E_{II}$& &	$E_B$ \\
\hline 
\multirow{3}{*}{$^{239}$Pu ($5/2^+)$}&	14&&	8.14& &	2.42& &	11.25 \\
				     &	16&&	7.97& &	2.22& &	10.83 \\
				     &	18&&	7.93& &	2.12& &	10.80 \\
\hline
\multirow{3}{*}{$^{240}$Pu}&	14&&	8.18& &	2.53& &	10.18	 \\
		 	   &	16&&	8.00& &	2.31& &	9.76	 \\
		 	   &	18&&	7.96& &	2.22& &	9.71	 \\
\end{tabular}
\end{ruledtabular}
\end{table}

\begin{acknowledgments}
M.H.K would like to thank Universiti Teknologi Malaysia for the
financial support throughout his doctoral study. Another
T.V.N.H. acknowledges the support by the U.S. NSF Grant No. 1415656
and U.S. DOE Grant No. DE-FG02-08ER41533. P.Q. gratefully acknowledges
the support of Dr. Wan Saridan for his stay in UTM during the
preparation of this paper using his research grant GUP No. 4F501 and
of the SCAC of the French Embassy in Kuala Lumpur.
\end{acknowledgments}

\bibliography{References}

\end{document}